\documentstyle[12pt,epsfig]{article}

\catcode`\@=11
\def\@citex[#1]#2{%
\if@filesw \immediate \write \@auxout {\string \citation {#2}}\fi
\@tempcntb\m@ne \let\@h@ld\relax \def\@citea{}%
\@cite{%
  \@for \@citeb:=#2\do {%
    \@ifundefined {b@\@citeb}%
      {\@h@ld\@citea\@tempcntb\m@ne{\bf ?}%
      \@warning {Citation `\@citeb ' on page \thepage \space undefined}}%
      {\@tempcnta\@tempcntb \advance\@tempcnta\@ne%
      \@tempcntb\number\csname b@\@citeb \endcsname \relax%
      \ifnum\@tempcnta=\@tempcntb 
	\ifx\@h@ld\relax%
	  \edef \@h@ld{\@citea\csname b@\@citeb\endcsname}%
	\else%
	  \edef\@h@ld{\ifmmode{-}\else--\fi\csname b@\@citeb\endcsname}%
	\fi%
      \else
	\@h@ld\@citea\csname b@\@citeb \endcsname%
	\let\@h@ld\relax%
      \fi}%
    \def\@citea{,\penalty\@highpenalty\,}%
  }\@h@ld
}{#1}}

\def\@citeb#1#2{{[#1]\if@tempswa , #2\fi}}
\def\@citeu#1#2{{$^{#1}$\if@tempswa , #2\fi }}
\def\@citep#1#2{{#1\if@tempswa , #2\fi}}

\def\bcites{         
	\catcode`\@=11
	\let\@cite=\@citeb
	\catcode`\@=12
}

\def\upcites{         
	\catcode`\@=11
	\let\@cite=\@citeu
	\catcode`\@=12
}

\def\plaincites{      
	\catcode`\@=11
	\let\@cite=\@citep
	\catcode`\@=12
}

\newcount\hour
\newcount\minute
\newtoks\amorpm
\hour=\time\divide\hour by 60
\minute=\time{\multiply\hour by 60 \global\advance\minute by-\hour}
\edef\standardtime{{\ifnum\hour<12 \global\amorpm={am}%
	\else\global\amorpm={pm}\advance\hour by-12 \fi
	\ifnum\hour=0 \hour=12 \fi
	\number\hour:\ifnum\minute<10 0\fi\number\minute\the\amorpm}}
\edef\militarytime{\number\hour:\ifnum\minute<10 0\fi\number\minute}

\def\draftlabel#1{{\@bsphack\if@filesw {\let\thepage\relax
   \xdef\@gtempa{\write\@auxout{\string
      \newlabel{#1}{{\@currentlabel}{\thepage}}}}}\@gtempa
   \if@nobreak \ifvmode\nobreak\fi\fi\fi\@esphack}
	\gdef\@eqnlabel{#1}}
\def\@eqnlabel{}
\def\@vacuum{}
\def\marginnote#1{}
\def\draftmarginnote#1{\marginpar{\raggedright\scriptsize\tt#1}}
\def\draft{
	\pagestyle{plain}
	\overfullrule=2pt
	\oddsidemargin -.5truein
	\def\@oddhead{\sl \phantom{\today\quad\militarytime} \hfil
	\smash{\Large\sl DRAFT} \hfil \today\quad\militarytime}
	\let\@evenhead\@oddhead
	\let\label=\draftlabel
	\let\marginnote=\draftmarginnote
	\def\ps@empty{\let\@mkboth\@gobbletwo
	\def\@oddfoot{\hfil \smash{\Large\sl DRAFT} \hfil}
	\let\@evenfoot\@oddhead}
	\def\@eqnnum{(\theequation)\rlap{\kern\marginparsep\tt\@eqnlabel}%
	\global\let\@eqnlabel\@vacuum}  }

\def\blackfonts{
	\font\blackboard=msbm10 scaled\magstep1
	\font\blackboards=msbm8
	\font\blackboardss=msbm6
}

\def\nblack{            
	\def\ZZ{{Z \n{10} Z}}
	\def\NN{{N \n{14} N}}
	\def\CC{{C \n{11} C}}
	\def\RR{{R \n{11} R}}
	\def\QQ{{Q \n{12} Q}}
	\def\PP{{P \n{11} P}}
}

\def\prep{         
	\catcode`\@=11
	\input art10.sty
	\catcode`\@=12
	
	\let\small\null
	\def\blackfonts{
		\font\blackboard=msbm10
		\font\blackboards=msbm7
		\font\blackboardss=msbm5
	}
	\let\sl\it
	\twocolumn
	\sloppy
	\voffset=-2.54truecm
	\hoffset=-2.54truecm
	\flushbottom
	\parindent 1em
	\leftmargini 2em
	\leftmarginv .5em
	\leftmarginvi .5em
	\marginparwidth 48pt
	\marginparsep 10pt
	\setlength{\columnsep}{2truecm}
	\setlength{\textwidth}{25.4truecm}
	\setlength{\textheight}{17truecm}
	\baselineskip=16pt
	\oddsidemargin .18truein
	\evensidemargin .17truein
}

\def\eqalign#1{\null\,\vcenter{\openup\jot\m@th
  \ialign{\strut\hfil$\displaystyle{##}$&$\displaystyle{{}##}$\hfil
      \crcr#1\crcr}}\,}
\def\eqalignno#1{\displ@y \tabskip\centering
  \halign to\displaywidth{\hfil$\@lign\displaystyle{##}$\tabskip\z@skip
    &$\@lign\displaystyle{{}##}$\hfil\tabskip\centering
    &\llap{$\@lign##$}\tabskip\z@skip\crcr
    #1\crcr}}

\def\section{\@startsection {section}{1}{\z@}{3.ex plus 1ex minus
 .2ex}{2.ex plus .2ex}{\large\bf}}
\def\subsection{\@startsection{subsection}{2}{\z@}{2.75ex plus 1ex minus
 .2ex}{1.5ex plus .2ex}{\bf}}

\def\appendix{{\newpage\section*{Appendix}}\let\appendix\section%
	{\setcounter{section}{0}
	\gdef\thesection{\Alph{section}}}\section}

\def\abstract{\if@twocolumn
\section*{Abstract}
\else 
\begin{center}
{\bf Abstract\vspace{-.5em}\vspace{0pt}}
\end{center}
\quotation
\fi}

\catcode`\@=12

\newcommand{\beq}{\begin{equation}}
\newcommand{\eeq}{\end{equation}}
\newcommand{\beqa}{\begin{eqnarray}}
\newcommand{\eeqa}{\end{eqnarray}}

\newcommand{\Z}{{\bf Z}}

\newcommand{\R}{{\bf R}}
\newcommand{\C}{{\bf C}}
\newcommand{\e}{{\rm e}}

\newcommand{\tilQ}{\widetilde{Q}}
\newcommand{\Qn}{Q_0}
\newcommand{\tilQn}{\widetilde{Q}_0}
\newcommand{\M}{{\cal M}}
\newcommand{\MV}{{\cal M}_V}
\newcommand{\bM}{\overline{\cal M}}

\newcommand{\HH}{{\cal H}}
\newcommand{\z}{\zeta}
\newcommand{\tild}{\widetilde{d}}

\def\noj#1,#2,{{\bf #1} (19#2)\ }
\def\jou#1,#2,#3,{{\sl #1\/ }{\bf #2} (19#3)\ }
\def\ann#1,#2,{{\sl Ann.\ Physics\/ }{\bf #1} (19#2)\ }
\def\cmp#1,#2,{{\sl Comm.\ Math.\ Phys.\/ }{\bf #1} (19#2)\ }
\def\ma#1,#2,{{\sl Math.\ Ann.\/ }{\bf #1} (19#2)\ }
\def\jd#1,#2,{{\sl J.\ Diff.\ Geom.\/ }{\bf #1} (19#2)\ }
\def\invm#1,#2,{{\sl Invent.\ Math.\/ }{\bf #1} (19#2)\ }
\def\cq#1,#2,{{\sl Class.\ Quantum Grav.\/ }{\bf #1} (19#2)\ }
\def\cqg#1,#2,{{\sl Class.\ Quantum Grav.\/ }{\bf #1} (19#2)\ }
\def\ijmp#1,#2,{{\sl Int.\ J.\ Mod.\ Phys.\/ }{\bf A#1} (19#2)\ }
\def\jmphy#1,#2,{{\sl J.\ Geom.\ Phys.\/ }{\bf #1} (19#2)\ }
\def\jams#1,#2,{{\sl J.\ Amer.\ Math.\ Soc.\/ }{\bf #1} (19#2)\ }
\def\grg#1,#2,{{\sl Gen.\ Rel.\ Grav.\/ }{\bf #1} (19#2)\ }
\def\mpl#1,#2,{{\sl Mod.\ Phys.\ Lett.\/ }{\bf A#1} (19#2)\ }
\def\nc#1,#2,{{\sl Nuovo Cim.\/ }{\bf #1} (19#2)\ }
\def\np#1,#2,{{\sl Nucl.\ Phys.\/ }{\bf B#1} (19#2)\ }
\def\pl#1,#2,{{\sl Phys.\ Lett.\/ }{\bf #1B} (19#2)\ }
\def\pla#1,#2,{{\sl Phys.\ Lett.\/ }{\bf #1A} (19#2)\ }
\def\pr#1,#2,{{\sl Phys.\ Rev.\/ }{\bf #1} (19#2)\ }
\def\prd#1,#2,{{\sl Phys.\ Rev.\/ }{\bf D#1} (19#2)\ }
\def\prl#1,#2,{{\sl Phys.\ Rev.\ Lett.\/ }{\bf #1} (19#2)\ }
\def\prp#1,#2,{{\sl Phys.\ Rept.\/ }{\bf #1C} (19#2)\ }
\def\ptp#1,#2,{{\sl Prog.\ Theor.\ Phys.\/ }{\bf #1} (19#2)\ }
\def\ptpsup#1,#2,{{\sl Prog.\ Theor.\ Phys.\/ Suppl.\/ }{\bf #1} (19#2)\ }
\def\rmp#1,#2,{{\sl Rev.\ Mod.\ Phys.\/ }{\bf #1} (19#2)\ }
\def\yadfiz#1,#2,#3[#4,#5]{{\sl Yad.\ Fiz.\/ }{\bf #1} (19#2) #3%
\ [{\sl Sov.\ J.\ Nucl.\ Phys.\/ }{\bf #4} (19#2) #5]}
\def\zh#1,#2,#3[#4,#5]{{\sl Zh.\ Exp.\ Theor.\ Fiz.\/ }{\bf #1} (19#2) #3%
\ [{\sl Sov.\ Phys.\ JETP\/ }{\bf #4} (19#2) #5]}

\def\beq{\begin{equation}}
\def\eeq{\end{equation}}
\def\beqar{\begin{eqnarray}}
\def\eeqar{\end{eqnarray}}
\def\non{\nonumber}

\newcommand{\be}{\begin{equation}}
\newcommand{\ee}{\end{equation}}
\newcommand{\bea}{\begin{eqnarray}}
\newcommand{\eea}{\end{eqnarray}}

\def\nfrac#1#2{{\displaystyle{\vphantom1\smash{\lower.5ex\hbox{\small$#1$}}%
	\over\vphantom1\smash{\raise.25ex\hbox{\small$#2$}}}}}

\def\p#1{\mskip#1mu}
\def\n#1{\mskip-#1mu}
\def\stop{\p6.}
\def\comma{\p6,}

\def\lae{\mathrel{\mathop{\smash{\lower .5 ex \hbox{$\stackrel<\sim$}}}}}
\def\lae{\mathrel{\mathop{\smash{\lower .5 ex \hbox{$\stackrel>\sim$}}}}}

\def\pa{\partial}

\def\l:{\mathopen{:}\,}
\def\r:{\,\mathclose{:}}

\def\vr{\vec{r}}

\def\vm{\vec{m}}
\def\vom{\vec{\omega}}
\def\vp{\vec{\phi}}

\catcode`\@=11
\def\theequation{\arabic{equation}}
\catcode`\@=12

\nblack
\bcites

\nblack

\catcode`\@=11
\def\theequation{\thesection.\arabic{equation}}
\@addtoreset{equation}{section}
\@addtoreset{footnote}{section}
\@addtoreset{footnote}{subsection}
\catcode`\@=12

\newcommand{\beqn}{\begin{equation}}
\newcommand{\eeqn}{\end{equation}}
\newcommand{\beqnarray}{\begin{eqnarray}}
\newcommand{\eeqnarray}{\end{eqnarray}}
\newcommand {\bear} [1] {\begin {array} {#1}}
\newcommand {\ear} {\end {array}}

\newcommand {\beqarn} {\begin{eqnarray*}}
\newcommand {\eeqarn} {\end{eqnarray*}}

\begin{document}
\begin{titlepage}

\begin{center}
\today
\hfill LBNL-039543, UCB-PTH-96/47 \\
\hfill                  hep-th/9611063

\vskip 1.5 cm
{\large \bf Mirror Symmetry in Three-Dimensional 
Gauge Theories, Quivers and D-branes
\\}
\vskip 1 cm 
{Jan de Boer, Kentaro Hori, Hirosi Ooguri and Yaron Oz}\\
\vskip 0.5cm
{\sl Department of Physics,
University of California at Berkeley\\
366 Le\thinspace Conte Hall, Berkeley, CA 94720-7300, U.S.A.\\
and\\
Theoretical Physics Group, Mail Stop 50A--5101\\
Ernest Orlando Lawrence Berkeley National Laboratory, 
Berkeley, CA 94720, U.S.A.\\}

\end{center}

\vskip 0.5 cm
\begin{abstract}
We construct and analyze 
dual $N=4$ supersymmetric gauge theories in three dimensions
with unitary and symplectic gauge groups.
The gauge groups and the field content of the theories
are encoded in quiver diagrams.
The duality exchanges the Coulomb and Higgs branches and the Fayet-Iliopoulos
and mass parameters.
We analyze the classical and the quantum moduli spaces of the theories and construct
an explicit mirror map between the mass parameters and the the Fayet-Iliopoulos
parameters of the dual.
The results generalize the relation between ALE spaces and moduli spaces
of $SU(n)$ and $SO(2n)$ instantons. 
We interpret some of these results from the
string theory viewpoint, for $SU(n)$ by analyzing 
T-duality and extremal transitions in type II string compactifications, 
for $SO(2n)$ by using D-branes as probes. Finally, we make a 
proposal for the moduli space of vacua of these theories in
the absence of matter.


\end{abstract}

\end{titlepage}

\section{Introduction}
$N=4$ supersymmetric gauge theories in three dimensions have been studied
recently from string theory as well as field theory viewpoints \cite{s,sw,si,ch}.
In these theories both the Coulomb and the Higgs branches are hyperk\"ahler 
manifolds.
In \cite{si} a duality between $N=4$ supersymmetric gauge theories 
in three dimensions has been proposed under which the Higgs and Coulomb branches
and the Fayet-Iliopoulos (FI) and mass terms are exchanged.
The dual gauge theories have an ALE space as Higgs branch, and were
based on Kronheimer's construction \cite{k}
of ALE spaces as an hyperk\"ahler quotient.

In this paper we generalize the duality (mirror) proposal to other 
$N=4$ supersymmetric gauge theories in three dimensions.
A gauge theory and its conjectured
dual will be called A-model and B-model respectively.
The gauge groups and field content of the theories are encoded in quiver diagrams
that correspond to Kronheimer-Nakajima's
hyperk\"ahler quotient construction of
quiver varieties \cite{kn,Nakajima2}, which will then automatically be the Higgs
branch of the associated gauge theory.
Specifically, we propose and study the duality between the following families of 
$N=4$ supersymmetric gauge theories:

\noindent 
(1) The A-model has $U(k)$ gauge group, $n$ hypermultiplets
in the fundamental representation of the gauge group 
and one hypermultiplet in the adjoint representation. Its dual B-model 
has $U(k)^n$ gauge group and matter content specified
by a quiver diagram corresponding to the Hilbert scheme of $k$ points on an ALE space
of $A_{n-1}$ type \footnote{The Hilbert schemes of $k$ points on complex surfaces have recently
appeared as the moduli spaces of D-branes \cite{HM,FKN}}.
By the Hilbert scheme of $k$ points on a complex surface $X$ we mean 
a smooth resolution
of the $k$-symmetric product of $X$, $Sym^k X$. Concretely,
there will be one hypermultiplet in the fundamental representation of one of the 
$U(k)$'s, and $n$ hypermultiplets 
charged under a pair of $U(k)$'s.

\noindent 
(2) The A-model has $Sp(k)$ gauge group, $n$ hypermultiplets
in the fundamental representation of the gauge group 
and one hypermultiplet in the antisymmetric representation. Its dual B-model 
has $U(k)^4U(2k)^{n-3}$ gauge group and matter content specified
by a quiver diagram corresponding to the Hilbert scheme of $k$ points on ALE space
of $D_{n}$ type.

\noindent
(3) The A and B models have $U(k)^n$ and $U(k)^m$ gauge groups respectively,
and matter content specified
by quiver diagrams corresponding to the hyperk\"ahler quotient construction of
certain moduli spaces of instantons on vector bundles over an ALE space
of $A_{n-1}$ type. This is a generalization of (1).

The paper is organized as follows:
Section 2 is a brief introduction to $N=4$ supersymmetric gauge theories in three dimensions.
In section 3 we define the dual gauge theories  
associated  with quiver diagrams. We present the proposed dualities, the
Higgs and Coulomb branches of the theories and the mirror map between the mass and
FI parameters.
In section 4 we study the first proposed family of dualities for $U(k)$ gauge theories.
We start by providing the first evidence to this duality proposal by counting the dimensions
of the Higgs and Coulomb branches as well as the number of mass  and FI terms.
We then study how the quantum corrections to the metric on the Coulomb branch 
fit into the mirror picture.
We compute the one-loop corrections to the hyperk\"ahler metric on the Coulomb
branch of the A-model and compare to the exact metric on the Higgs branch of the
B-model. The comparison yields strong support for the
 mirror map between the mass terms of the A-model
and the FI terms of the  B-model.
In section 5 we analyze the structure of  the Coulomb, 
Higgs and mixed branches 
for various mass and FI parameters. We observe a complete agreement
of their dimensions which provide further evidence for the duality.
In particular, we complete the proof of the mirror map by fixing the
ambiguities left after the one-loop computation. 
We show how the proposed duality completely determines 
the quantum moduli space of vacua.
In section 6 we examine type II string compactifications that in the
field theory limit yield the A-model. The gauge symmetry and matter fields
arise by wrapping D-branes around vanishing cycles and we use
T-duality and 
 extremal
transitions to explain the gauge theory duality from a stringy
viewpoint.
In section 7 we study the second proposed family of dualities for $Sp(k)$ gauge theories.
We  provide the counting  evidence for this duality proposal,
study  the quantum corrections,
 derive the mirror map and use D-brane probes and the Type I - M-theory duality to further
support the gauge theory picture. 
In section 8 we 
study the third proposed family of dualities for $U(k)^n$ gauge theories.
We  provide the counting  evidence for this duality proposal,
study the Higgs, Coulomb and mixed  branches of the dual theories, and
give the mirror map.
In section 9 we discuss the case of $U(k)$, $SU(k)$ and $Sp(k)$ gauge theories
without matter, present a proposal for their moduli spaces, and conclude with
open problems.

\section{$N=4$ supersymmetric gauge theories in three dimensions}

We begin with a brief review of $N=4$ supersymmetric gauge theories in three
dimensions.

$N=4$ supersymmetric gauge theories in three
dimensions can be constructed by dimensional reduction of $N=1$ supersymmetric
gauge theories in six dimensions. The R-symmetry group is
$SU(2)_L\times SU(2)_R$ with 
$SU(2)_L$ being the double cover of rotations in the three reduced coordinates
and $SU(2)_R$ is the R symmetry group in six dimensions.
The masses and FI parameters transform under $SU(2)_L\times SU(2)_R$ 
as $({\bf 3,1})$ and $({\bf 1,3})$ respectively. The mass terms deform the metric
 on the Coulomb branch and lift some of the Higgs branch,
  while the FI terms deform the metric on the Higgs branch and
  lift some of the Coulomb branch.
 The Higgs branch is constructed as a hyperk\"ahler quotient 
 with an $SU(2)_R$ action,
 and unlike
 the Coulomb branch is not modified
 by quantum corrections.
 
 The $N=4$ vector multiplet in three dimensions contains three scalars
 $\phi^{\alpha}, \alpha=1,2,3$ which transform as $({\bf 3,1})$
 under the R-symmetry group.
 Their potential energy is
 \beq
 V = \frac{1}{e^2} \sum_{\alpha < \beta}Tr[\phi^{\alpha},\phi^{\beta}]^2
 \comma
 \label{V}
 \eeq
 where $e$ is the gauge coupling.
 The potential energy vanishes if the $\phi^{\alpha}$ commute and thus they take
 values in a common Cartan sub-algebra of the gauge group. 
 For a generic vev 
 in this Cartan subalgebra,
 the gauge group of rank $r$ is
 broken to $U(1)^r$. Thus, in addition to the $3r$ scalars we have 
 $r$ massless photons which are dual to $r$ scalars in  three dimensions.
 The Coulomb branch is parametrized by the vevs of the $3r$ scalars and the $r$
 scalars dual to the photons and thus is of dimension $4r$.
 Due to the $N=4$ supersymmetry it is a  hyperk\"ahler manifold with
 an $SU(2)_L$ action.
 Its metric is corrected by loop and monopole corrections.
 The monopoles are instantons in three dimensions and they provide exponential corrections
 to the metric on the Coulomb branch.
 
 The duality between $N=4$ supersymmetric gauge theories 
in three dimensions exchanges the  Higgs and Coulomb branches,
 the Fayet-Iliopoulos (FI) parameters and masses and the R-symmetry groups $SU(2)_L$ and
 $SU(2)_R$. The fact that the Higgs branch is not modified by quantum corrections while the Coulomb
 branch is, implies that like in mirror symmetry in string theory
  quantum corrections in one model are seen at the classical level of the
 dual and vice versa.
 Note that in general the duality between the A-models and B-models becomes exact
 only when the bare coupling constant $e^2$ is sent to infinity.

\section{Mirror symmetric gauge theories and quivers}

 In this section we define the gauge theories  
associated  with quiver diagrams. We present the proposed dualities, the
Higgs and Coulomb branches of the theories and the mirror map between the mass and
FI parameters.
An object that will appear frequently in the discussion is the Hilbert scheme of $k$ points on a
complex surface $X$, $Hilb^{[k]} X$. As we noted previously, this is a resolution 
of the the quotient singularities\footnote{We use the terminology
 quotient singularity
to denote the singularities that arise in a symmetric product due
to the action of the symmetric group.} of the $k$-symmetric product of $X$, $Sym^kX$.
In the A model the Coulomb branch will be
described by a Hilbert scheme
and the parameter for the 
resolution of the quotient singularities will be found to be
 the adjoint hypermultiplet
mass $\vm_{adj}$ for $U(k)$ gauge theories and the mass of the antisymmetric hypermultiplet
$\vm_{as}$ for the $Sp(k)$ gauge theories.
The parameters for the resolution of the singularities of the complex surface $X$ will be shown
to correspond to the
masses of the fundamental hypermultiplets $\vm_{fund}$ in both cases.
In the B-model the Higgs branch 
will be described by a Hilbert scheme and 
the parameters for the resolution of
all the singularities will be explicitly constructed from the FI
parameters.

\subsection{$U(k)$ Gauge Groups}

\addtocounter{footnote}{1}

The A-model has a $U(k)$ gauge group,  $n$ 
hypermultiplets
in the fundamental representation
and one hypermultiplet in the adjoint representation.
This is precisely the field content needed for the 
hyperk\"ahler quotient
construction of the moduli space of $SU(n)$ 
$k-$instantons $\bM_k(SU(n))$ 
\cite{Donaldson},\footnote{By
$\bM_k(SU(n))$ we denote an enlarged moduli space 
which includes the small instantons.
For more technical details see section~5.}
which is indeed
the Higgs branch of the A-model.

The B-model is associated with the quiver diagram in 
figure 1.

\begin{figure}
\begin{center}
$\mbox{\epsfig{figure=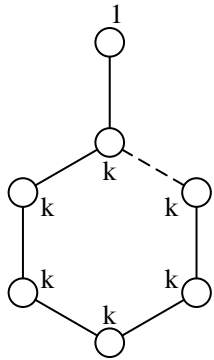}}$
\end{center}
\caption{Quiver diagram for the B-model of $U(k)$ gauge 
theory}
\end{figure}

We attach an index $k_i$ at each node $i$. There
are $n$ nodes in the diagram with $k_i=k$ and one node with $k_i=1$.
The  gauge group and the field content of the theory are 
encoded in the diagram in
the following way:
We associate to each node $i$ with $k_i=k$
a gauge group $U(k)_i$,
to each link
${}_i\!\circ\!\!-\!\!\!-\!\!\!-\!\circ_j$
with $k_i=k_j=k$ a hypermultiplet in the
representation $({\bf k},{\bf k^*})$ of $U(k)_i\times 
U(k)_j$,
and to the link attached to the node with index 1
a hypermultiplet in the fundamental representation of 
the $U(k)$
gauge group associated with the other node of the link.
This is the field content needed for the hyperk\"ahler 
quotient
construction of the
Hilbert scheme of $k$ points on ALE space
of type $A_{n-1}$, which we will denote by
$X_{A_{n-1}}$\cite{kn,Nakajima2},
and which is the Higgs branch of the B-model.
The duality between the moduli spaces  can be roughly 
summarized
by the following table:
\vskip 0.2cm
\begin{center}
\renewcommand{\arraystretch}{1.5}
\begin{tabular}{|c|c|c|}     \hline
Model  & ${\cal M}_V$ & ${\cal M_H}$ \\ \hline
       A & $Hilb^{[k]} X_{A_{n-1}}$ & $\bM_k(SU(n))$ \\  
\hline     
       B &  $\bM_k(SU(n))$ &  $Hilb^{[k]} X_{A_{n-1}}$ 
                            \\ \hline 
\end{tabular}
\end{center}
\begin{center}
Table 1: The Coulomb and Higgs branches of A and B 
models
\end{center}
The precise structure is more detailed and depends on 
the mass and FI parameters.
Consider the A-model:
Without mass terms, the vector multiplet moduli space is 
the $k$-symmetric product $Sym^kX_{A_{n-1}}$
of the ALE space. It has singularities inherited from 
the simple singularity of $A_{n-1}$ type of 
$X_{A_{n-1}}$,
and also singularities coming from 
modding out by the action of the
symmetric 
group.
The masses for the fundamental hypermultiplets resolve 
the simple singularity of $X_{A_{n-1}}$. We denote the 
resolved ALE space as $\widetilde{X}_{A_{n-1}}$.
The mass of the adjoint hypermultiplets resolves 
the quotient singularities of the symmetric product.
In the following table, we show how the vector multiplet 
moduli space depends on the mass parameters\footnote{In
fact, there are two independent mass parameters 
$\vm_{U(1)}$ and $\vm_{SU(k)}$
for the adjoint hypermultiplet, associated to its trace and 
traceless part respectively. The metric on the A-model moduli
space does not depend on $\vm_{U(1)}$. Its only effect
is to lift a trivial direction in the Higgs branch, corresponding
to the center of mass of the instantons.
Consequently, we do not count $\vm_{U(1)}$ as an
independent parameter, and there is no corresponding
FI parameter in the B-model.
Our duality applies to the cases 
where $\vm_{SU(k)}$ and $\vm_{U(1)}$ are either
both vanishing or both non-vanishing.}.
\begin{center}
\renewcommand{\arraystretch}{1.5}
\begin{tabular}{|c|c|}     \hline
Masses & ${\cal M}_V$  \\ \hline
 $\vm_{fund}=0$, $\vm_{adj}=0$ & $Sym^k X_{A_{n-1}}$ \\ 
\hline
 $\vm_{fund}\neq 0$, $\vm_{adj}=0$ & $Sym^k 
\widetilde{X}_{A_{n-1}}$
  \\ \hline
  $\vm_{fund}=0$, $\vm_{adj}\neq 0$ & $Hilb^{[k]} 
X_{A_{n-1}}$  \\ \hline
  $\vm_{fund}\neq 0$, $\vm_{adj}\neq 0$ & 
$\,\,\,Hilb^{[k]}\widetilde{X}_{A_{n-1}}\,\,\,$  \\ 
\hline                                                           
\end{tabular}
\end{center}
\begin{center}
Table 2: Mass parameters versus the vector multiplet moduli space 
(A-model)
\end{center}
The other  effect of the mass terms is to lift some of the flat 
directions of the hypermultiplet moduli space. In section 5 we will analyze how
this lifting is compatible with the resolution of the singularity.
 
In the B-model, the resolution of the singularity of the
hypermultiplet moduli space
and  the lifting  of some of the flat directions for the vector 
multiplets
are caused by turning on FI terms. The  way in which the moduli 
spaces
are resolved or lifted  matches exactly with the 
A-model
when the vector multiplets and hypermultiplets are
exchanged,
provided that the FI parameters are related to the mass 
parameters of the A-model.

 The mirror map between the mass parameters
of the A-model and the FI parameters of the B-model 
takes the form
\beq
\vec{m}_i = \sum_{l=0}^i \vec{\zeta}_l,~~~~~~~~~~
\vec{m}_{adj} = \sum_{l=0}^{n-1} \vec{\zeta}_l
\comma
\label{map}
\eeq
where $\vec{m}_i$ are the masses of the fundamental 
hypermultiplets,
$\vec{m}_{adj}$ is the mass of the adjoint 
hypermultiplets and
$\vec{\zeta}_l$ are the FI parameters.
 Note that a linear combination
of masses can be eliminated for every $U(1)$ factor in 
the gauge group by
shifting the origin of the Coulomb branch. In 
(\ref{map})
we used this freedom to choose 
$\vec{m}_{n-1}=\vec{m}_{adj}$.

The first evidence that we will provide
for the duality between the A and B models will be the 
matching  of 
the dimensions of the Higgs and Coulomb branches and the 
number of FI and mass
terms. We will then analyze the one-loop corrections and
derive the mirror map (\ref{map}).
A detailed analysis of the moduli spaces will provide 
further evidence for the duality, which will 
in particular completely determine the mirror map,
fixing all remaining ambiguities.
Finally we will show how the duality structure arises 
from a stringy D-brane picture.

\subsection{$Sp(k)$ Gauge Groups}

We define the A-model to have $Sp(k)$ as its gauge group.
The matter content consists of  $n$ hypermultiplets
in the fundamental representation of $Sp(k)$  
and one hypermultiplet in the antisymmetric representation of $Sp(k)$.
The Higgs branch of the A-model is the moduli space of $SO(2n)$ 
$k-$instantons $\bM_k(SO(2n))$\cite{wein}.

The B-model is associated with the quiver diagram in figure 2.

\begin{figure}
\begin{center}
$\mbox{\epsfig{figure=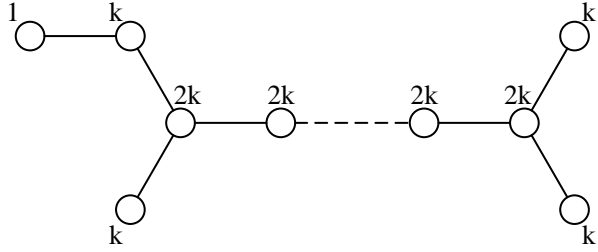}}$
\end{center}
\caption{Quiver diagram for the B-model of $Sp(k)$ gauge theory}
\end{figure}

As described in the previous section
we associate to each node a gauge group corresponding to its index.
Diagram 2 has four nodes with index $k$ and $n-3$ nodes with index $2k$,
 thus the gauge group
of the B-model is $U(k)^4U(2k)^{n-3}$. Again,  the
the matter content is $\oplus_{ij}
a_{ij}({\bf k_i},{\bf k_j}^{\ast})$ where $a_{ij}$ is one if there is a link between
the nodes $i$ and $j$ and zero otherwise. In addition,
there is one fundamental 
hypermultiplet
charged with respect to the $U(k)$
associated to the node that is connected to the exceptional one.
The Higgs branch of the B-model is the
Hilbert scheme of $k$ points on an ALE space
of $D_n$ type \cite{kn}.

The duality is roughly summarized in the following table:
\vskip 0.2cm
\begin{center}
\renewcommand{\arraystretch}{1.5}
\begin{tabular}{|c|c|c|}     \hline
Model   & ${\cal M}_V$ & ${\cal M_H}$   \\ \hline
       A & $Hilb^{[k]} X_{D_{n}}$ & $\bM_k(SO(2n))$  \\  \hline     
       B & $\bM_k(SO(2n))$ & $Hilb^{[k]} X_{D_{n}}$\\ \hline 
\end{tabular}
\end{center}
\begin{center}
Table 3: The Coulomb and Higgs branches of A and B models
\end{center}

As for the $U(k)$ case, the detailed structure depends on the 
mass and FI parameters.
An illustrative table for the effect of the mass 
parameters\footnote{Here,
$\vm_{as}$ denotes the mass parameter for the
hypermultiplet in the anti-symmetric representation.
As in the $U(k)$ case, there are really two
mass parameters for the anti-symmetric representation,
one of which corresponds to the trivial representation,
and the same statements made in the footnote
for $U(k)$ apply here too.}
 on the Coulomb branch is 

\vskip 0.2cm
\begin{center}
\renewcommand{\arraystretch}{1.5}
\begin{tabular}{|c|c|}     \hline
Masses  & ${\cal M}_V$ \\ \hline
 $\vm_{fund}=0$, $\vm_{as}=0$ & $Sym^k X_{D_{n}}$ \\ \hline
 $\vm_{fund}\neq 0$, $\vm_{as}=0$ & $Sym^k \widetilde{X}_{D_{n}}$
   \\ \hline
  $\vm_{fund}=0$, $\vm_{as}\neq 0$ & $Hilb^{[k]} X_{D_{n}}$  \\ \hline
  $\vm_{fund}\neq 0$, $\vm_{as}\neq 0$ & \,\, $Hilb^{[k]}\widetilde{X}_{D_{n}}$ 
  \,\,  \\ \hline                                                         
\end{tabular}
\end{center}
\begin{center}
Table 4: 
Mass parameters versus the vector multiplet moduli space (A-model)
\end{center}

The structure of the moduli space of the 
B-model can be read of by exchanging the 
Higgs and Coulomb branches of the A-model.
The masses of the the hypermultiplets in the fundamental
representation $\vm_i$ and the antisymmetric hypermultiplet mass $\vm_{as}$ of the A-model 
are mapped under the duality
to the FI parameters of the B-model
\beqar
\vec{m}_{i} &=& 2\sum_{l=1}^{i} \vec{\zeta}_l + \vec{\zeta}_{n-1} + \vec{\zeta}_n~~~~~~
i < n,~~~~~~~~\vec{m}_{n} = \vec{\zeta}_n - \vec{\zeta}_{n-1}\comma \nonumber\\
\vec{m}_{as} &=& \vec{\zeta}_0 + \vec{\zeta}_1 + 2\sum_{l=2}^{n-2} \vec{\zeta}_l
+ \vec{\zeta}_{n-1} + \vec{\zeta}_n.
\label{mapas}
\stop
\eeqar   
Here, $\vec{\zeta}_0$ is associated to the node connected to the exceptional one, 
$\vec{\zeta}_1$ to the other leftmost node with index $k$, $\vec{\zeta}_l$ for
$1<l<n-1$ to the nodes with index $2k$ ordered from left to right, and
$\vec{\zeta}_{n-1}$ and $\vec{\zeta}_n$ to the rightmost nodes with index $k$.

We will study this duality in section 7.
We will provide the counting evidence, analyze the quantum corrections,
derive the mirror map
and support the duality by a D-brane picture based on the Type I - M-theory
duality, and by the use of D-branes as probes.

\subsection{$U(k)^n$ Gauge Groups}
The gauge field
and matter content of the A and B models are encoded in the
quiver diagram in figure 3.

\begin{figure}
\begin{center}
$\mbox{\epsfig{figure=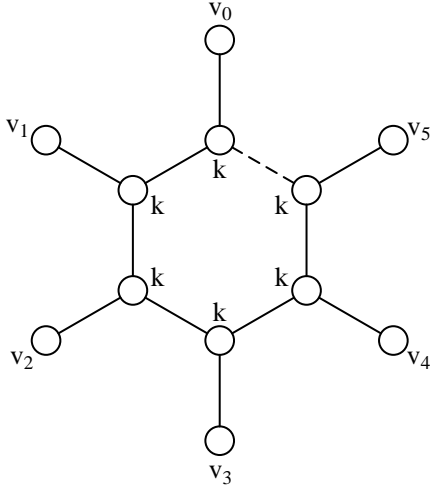}}$
\end{center}
\caption{Quiver diagram for the A-model of $U(k)^n$ gauge theory}
\end{figure}

The A-model gauge
group is $U(k)^n$, one $U(k)$ for each node of the extended Dynkin diagram.
Notice that there is no gauge symmetry associated to the outside nodes
with labels $v_i$. 
There  are two kinds of matter. As before,
for each pair of gauge groups whose nodes are
connected by an edge there will be matter transforming as $({\bf k},{\bf k}^{\ast})$ under 
$U(k)\times U(k)$. In addition, there will be $v_i$ matter fields transforming
in the fundamental representation of the $i$th $U(k)$ gauge group. 
We will denote the A-model as $(U(k)^n;\{v_i\})$. 
The Higgs branch of the A-model is  the moduli space of instantons
 on a vector bundle $V$ over an ALE space of type $A_{n-1}$. More precisely,
it describes the moduli space ${\cal M}_k(V)$
of instantons of instanton number $k$ on
 $V=\oplus {\cal R}_i^{\otimes v_i}$, with gauge group $U(V)$,
where ${\cal R}_i$ are particular line
bundles 
over the ALE space associated to the different representations of ${\bf Z}_n$
\cite{kn}. 
The B-model gauge theory is $(U(k)^m;\{w_i\})$, where
\beq
m=\sum_{i=0}^{n-1} v_i,~~~~~~~~~ n=\sum_{i=0}^{n-1} w_i
\stop
\eeq
The numbers $v_i$ and $w_i$ are
related as follows:
 Consider a Young diagram whose rows have lengths $\sum_{i=0}^p v_i$,
$p=0 ,\ldots ,n-1$. The lengths of the columns of this diagram can be parametrized as
$\sum_{i=0}^q w_i$, $q=0,\ldots m-1$, and the integers $w_i$ are the
ones appearing in the dual gauge theory. For example, $(U(k)^5,\{2,3,0,1,0\})$ is
proposed as the dual of $(U(k)^6,\{2,2,0,0,1,0\})$.
The $U(k)$ gauge theory we
considered so far in this paper corresponds to a Young diagram which is 
a rectangle of size $n \times 1$.  

The duality is summarized in the following table :
\vskip 0.2cm
\begin{center}
\renewcommand{\arraystretch}{1.5}
\begin{tabular}{|c|c|c|}     \hline
Model   & ${\cal M}_V$ & ${\cal M_H}$   \\ \hline
       A & ${\cal M}_k(\oplus {\cal R}_i^{\otimes w_i})$
       & ${\cal M}_k(\oplus {\cal R}_i^{\otimes v_i})$ \\  \hline     
       B & ${\cal M}_k(\oplus {\cal R}_i^{\otimes v_i})$
        & ${\cal M}_k(\oplus {\cal R}_i^{\otimes w_i})$\\ \hline 
\end{tabular}
\end{center}
\begin{center}
Table 5: The Coulomb and Higgs branches of A and B models
\end{center}

An important feature of this construction is that 
the dual of the dual theory is the
original theory again, as one would expect, making duality a true involution in
this  set of theories. 
In section~8, we will provide counting evidence for the duality and 
analyse the structure of the moduli spaces, and give arguments for the
following mirror map: Denote by $\vm_i^{(B)}$, $\sum_{l=0}^{j-1} w_l \leq i 
< \sum_{l=0}^j w_l$  the masses
of the hypermultiplets in the B-model charged only under the $j^{\rm th}$
$U(k)$. In addition, there are $m$ masses of hypermultiplets charged under
two $U(k)$'s. Using the freedom to shift the origin on the Coulomb branch,
we can choose all these masses equal to same value which we denote by
$\vm_{2f}^{(B)}$. This leaves only the freedom to add a constant
simultaneously to all $\vm_i^{(B)}$, which we use to fix $\vm_{n-1}^{(B)}=0$.
Then the relation between the FI parameters $\vec{\zeta}_i^{(A)}$ of the
A-model and the masses of the B-model reads
\bea
\sum_{l=0}^{i} \vec{\zeta}_l^{(A)} & = & \vm_i^{(B)} + (\sum_{l=0}^{i} v_l) \vm_{2f}^{(B)} 
\nonumber \\{}
\sum_{l=0}^{n-1} \vec{\zeta}_l^{(A)} & = & (\sum_{l=0}^{n-1} v_l) \vm_{2f}^{(B)}.
\label{3mirr}
\eea

\section{Duality for $U(k)$ Gauge Theories I: 
 Quantum Corrections and Mirror Map}

In this section we begin by providing the first preliminary
counting evidence for the duality.
We then turn to the computation of the one-loop corrections to the metric on the 
Coulomb branch of the $U(k)$ A model.
We further  compute the metric on the Higgs branch of
the B  model in the case where the sum of the Fayet-Iliopoulos terms vanishes.
This corresponds in the A model to the case where the mass
of the adjoint hypermultiplet vanishes. 
By comparing the two computations we derive the form of the mirror map
between the fundamental hypermultiplets mass parameters of A model and the 
FI parameters of B model for $\vm_{adj} =0$.
Finally we construct the mirror map with a non-vanishing  adjoint mass.

\subsection{Counting Evidence}

As a first evidence for the duality between the A and B models we count 
in quaternionic units 
the dimensions of the Higgs and Coulomb branches and the number of independent 
FI and mass terms. \\
\noindent\\
{\bf A-model:} The dimension of the Coulomb branch is the rank of the gauge group
$U(k)$
which is $d_V=k$. 
The Higgs branch is given by a hyperk\"ahler quotient construction and 
accordingly, its dimension equals
the dimension of the space of 
hypermultiplets minus the dimension of the gauge group.
Therefore, $d_H = (nk + k^2) - k^2 = nk$.
The number of FI terms is the number of $U(1)$ factors in the gauge group,
$n_{\zeta}=1$.
In order to count the number of mass parameters note that a linear combination
of masses can be eliminated for every $U(1)$ factor in the gauge group by
shifting the origin of the Coulomb branch. Thus, in this case
the number of mass parameters is $n_m = (n+1) -1 =n$.\\
\noindent\\
{\bf B-model:} The dimension of the Coulomb branch 
 is the rank of $U(k)^n$, thus $d_V=nk$. 
 The dimension of the Higgs branch is the dimension of the
space of hypermultiplets ($n k^2 + k$) 
minus the dimension of the gauge group ($nk^2$), thus  $d_H = k$.
The number of FI terms is the number of $U(1)$ factors in $U(k)^n$ 
and therefore
 $n_{\zeta}=n$.
The number of mass parameters is
$n_m = (n+1) -n =1$.
The counting is summarized in the following table:
\vskip 0.2cm
\begin{center}
\renewcommand{\arraystretch}{1.5}
\begin{tabular}{|c|c|c|c|c|}     \hline
Model   & $d_V$ & $d_H$ & $n_{\zeta}$ & $n_m$ \\ \hline
       A  & $k$ & $nk$ & $1$ & $n$ \\  \hline     
       B &  $nk$ & $k$ & $n$ & $1$ \\ \hline 
\end{tabular}
\end{center}
\begin{center}
Table 6: The dimension of the Coulomb and Higgs branches
and the number of mass and FI parameters of A and B models
\end{center}

The counting shows that we indeed have a symmetry under A-model 
$\leftrightarrow$ B-model,
$d_V \leftrightarrow d_H$ and $n_{\zeta} \leftrightarrow n_m$
which is a necessary condition for the duality to hold.

\subsection{A model - One-loop Corrections}
Consider the A model with gauge group $U(k)$, one hypermultiplet
in the adjoint representation and $n$ hypermultiplets in the fundamental
representation.
Let us parametrize the scalars that minimize the potential energy
(\ref{V}) by
\beq
\vp = diag[\vr_1,...,\vr_{k}]
\comma
\eeq
where $\vp = (\phi^1,\phi^2,\phi^3)$.

The one-loop corrected metric of the Coulomb branch of A model 
takes the form 
\beq
ds^2 = g_{ab}d\vr_a d\vr_b + (g^{-1})_{ab}(d\sigma_a + \vom_{ac}\cdot d\vr_c)
(d\sigma_b+ \vom_{bd}\cdot d\vr_d)
\comma
\label{jan1}
\eeq
where
\beqar
g_{aa} &=& \frac{1}{e^2} + \sum_{b\neq a}^{k}\left(
 \frac{-2}{|\vr_a-\vr_b|}
+ \frac{1}{|\vr_a-\vr_b +\vm_{adj}|}
+\frac{1}{|\vr_a-\vr_b -\vm_{adj}|} \right) +
\sum_{i=0}^{n-1}\frac{1}{|\vr_a-\vm_i|} \non\\
g_{ab} &=&\frac{2}{|\vr_a-\vr_b|}
- \frac{1}{|\vr_a-\vr_b +\vm_{adj}|}
- \frac{1}{|\vr_a-\vr_b -\vm_{adj}|}~~~~~~~~~a \neq b
\comma
\label{one}
\eeqar
where $a,b=1...k$.
The $\sigma_a$ are variables dual to the photons that remain massless
on the Coulomb branch. They are periodic with period $2\pi$, and constant
shifts of the $\sigma_i$ are triholomorphic isometries of the hyperk\"ahler
metric (\ref{one}). These isometries are unbroken in perturbation
theory, and every hyperk\"ahler metric of real dimension $4k$ with $k$
commuting triholomorphic isometries can be written in the form (\ref{jan1}),
where $g$ and $\omega$ satisfy \cite{roc1,roc2,PP}
\bea
\vec{\nabla}_a g_{bc} & = & \vec{\nabla}_b g_{ac} \nonumber \\{}
\frac{\partial}{\partial r_a^p} \omega^q_{bc} -
\frac{\partial}{\partial r_b^q} \omega^p_{ac} & = & \epsilon_{pqr}
\frac{\partial}{\partial r_a^r} g_{bc}.
\label{kahler}
\eea
This explains the form of the metric in (\ref{jan1}), and can be used
to express $\vom_{ab}$ in terms of $g_{ab}$. Thus,
in order to derive this form of the one-loop corrected metric we only need to 
look at  
the terms in the one-loop effective action coming from one-loop diagrams with 
two gauge fields on the external legs and the vector multiplet or hypermultiplet
running in the loop.
We then make use of the following limits:\\
\noindent
(1) Reduction in color: Taking the limit $|\vr_{k}| \rightarrow \infty$
is a reduction in the number of colors and we should recover the formula for 
the metric for the gauge group $U(k-1)$.
This implies that the coefficients of the different terms are independent
of the number of colors.
Thus it is sufficient to consider the gauge group $U(2)$.
The gauge group $U(1)$ is evidently not sufficient since the theory is free
in the absence of matter.
\\
\noindent
(2) Reduction in flavor: Taking the limit $|\vm_{n-1}| \rightarrow \infty$
is a reduction in the number of hypermultiplets in the fundamental
representation of the gauge group,
 and we should recover the formula for 
the metric for $n-1$ flavors.
This implies that the coefficients of the different terms are independent
of the number of flavors.\\
\noindent
(3) The first equation in (\ref{kahler}) implies
that the contributions of the vector multiplet and the adjoint
hypermultiplet to the diagonal and off diagonal elements of the metric are
of opposite sign and the same absolute value. It also implies that
 the hypermultiplets in the fundamental can contribute only to the diagonal
 terms of the metric.
 In order to see these it is sufficient to consider the $U(2)$ gauge group
 and use the equation for the metric $\pa_1g_{21} = \pa_2g_{11}$ implied by (\ref{kahler}).\\
 \noindent
(4) Reduction of the gauge group to $U(1)$ and considering the case of $n$
hypermultiplets in the fundamental representation while taking the limit
$|\vm_{adj}| \rightarrow \infty$, should recover for massive hypermultiplets
the Taub-NUT metric for a resolved $A_{n-1}$ singularity \cite{sw}
\beq
ds^2 = g^{-2}d\vr d\vr + g^{2}(d\sigma + \vom\cdot d\vr)^2
\comma
\eeq
where
\beq
g^{-2}(\vr) = \frac{1}{e^2} + 
\sum_{i=0}^{n-1}\frac{1}{|\vr-\vm_i|}
\comma
\eeq
and
\beq
\vec{\nabla}(g^{-2}) = \vec{\nabla}\times \vom
\stop
\eeq
This fixes the coefficient of the fundamental hypermultiplet contribution
to the metric.\\
\noindent
(5) Reduction of the gauge group to $SU(2)$ and considering the case of $n=2$
hypermultiplets in the fundamental representation while taking the limit
$|\vm_{adj}| \rightarrow \infty$, should recover for massless hypermultiplets
the classical metric since there are no quantum corrections in this case
\cite{sw}.
Using (4), this fixes the coefficient of the vector multiplet contribution
to the metric.
In order to see this explicitly consider the case of gauge group $U(2)$ with
two massless hypermultiplets in the fundamental representation.
For the metric $g_{ab}, a,b=1,2$ we take
\beqar
g_{aa} &=& \frac{1}{e^2} +\frac{\alpha}{|\vr_a-\vr_b|}
 +
\frac{2}{|\vr_a|} \non\\
g_{ab} &=&\frac{-\alpha}{|\vr_a-\vr_b|}
~~~~~~~~~a \neq b
\stop
\eeqar
where $\alpha$ is the constant coefficient to be determined and the coefficient 
of the fundamental hypermultiplets has been determined in (4).
Define
\beq
g_{ab}d\vr_a d\vr_b  = g_{++}(d\vr_+)^2 +
 2 g_{+-} d\vr_+ d\vr_- +
 g_{--}(d\vr_-)^2 
\comma
\eeq
where $\vr_{\pm} = \frac{\vr_1 \pm \vr_2}{\sqrt{2}}$.
$g_{++}$ and $g_{--}$ correspond to the $U(1)$ and $SU(2)$ parts of the metric
respectively. Restricting to the $SU(2)$ part, 
$\vr_1 = -\vr_2$,
 and requiring that $g_{--}$ does not get 
quantum corrections for two massless fundamentals
we get the required result $\alpha=-2$.\\
\noindent
(6) The coefficient of the adjoint hypermultiplet contribution is fixed
by reading from the 
Lagrangian its relation to that of the fundamental hypermultiplets.
Note that in the absence of hypermultiplets in the fundamental 
representation
there are no one-loop corrections to the metric if there is no adjoint mass.
This is consistent with the fact that in this case we have an $N=8$
 supersymmetry as a reduction of the $N=4$ supersymmetry in four dimensions.
 In this case the complex structure of the hyperk\"ahler manifold is, as expected 
 \cite{sw}, the same as that of the Jacobian corresponding to the $N=4$ curve
 \cite{sw2}. 

Consider the case  
with zero adjoint mass and $n$ massive
fundamentals, in the limit $e^2 \rightarrow \infty$.
In this case the one-loop metric describes the $k$-symmetric
product of resolved ALE spaces of $A_{n-1}$ type $\widetilde{X}_{A_{n-1}}$
(the symmetric product arises because we still have to divide by the
 action of the Weyl group $S_k$ of $U(k)$)
\beq
{\cal M}_V^{one-loop}({\rm A-model},\vm_{adj}=0, \vm_{fund}\neq 0)
  = Sym^k \widetilde{X}_{A_{n-1}}
\comma
\eeq 
where the masses of the hypermultiplets resolve the ALE singularities.
We will argue in the next section that this result is in fact exact, namely
\beq
{\cal M}_V^{Exact}({\rm A-model},\vm_{adj}=0, \vm_{fund}\neq 0)
  = Sym^k \widetilde{X}_{A_{n-1}}
\stop
\label{exact}
\eeq

When the adjoint mass is nonzero, $\vm_{adj}\neq 0$,
the one-loop metric is not positive definite
in the region $|\vec{r}_a - \vec{r}_b| \rightarrow 0$. 
We expect monopole corrections to contribute in this case, and that the metric
will become positive definite upon including these corrections. A similar 
phenomenon happens in pure $SU(2)$ gauge theory with zero or
one hypermultiplet in the fundamental representation \cite{sw},
and also when considering monopole
moduli spaces \cite{gm}. 
More specifically, in the region $|\vec{r}_a - \vec{r}_b| \ll |\vm_{adj}|$
for some $a,b$, while keeping other pairs $\gg |\vm_{adj}|$, the 
system can be well approximated by the $SU(2)$ gauge theory with
one adjoint hypermultiplet with bare mass $\vm_{adj}$. 
By a slight generalization of the results in \cite{sw} we see that
there are no higher-loop corrections in this region,
and we expect monopole corrections to restore the positivity of the
metric. 
There is a close analogy between the quotient singularity
$ \vec{r}_a \leftrightarrow \vec{r}_b$
in our case and the ${\bf Z}_2$ 
singularity 
$\vec{r} \rightarrow -
\vec{r} $  in the $SU(2)$ case, which is resolved by
monopole corrections. Since we expect monopole
corrections when $\vm_{adj}\neq 0$, this
suggests that the adjoint mass is a parameter for the
resolution of the quotient singularities of the symmetric product. 
In the following sections we will provide further support to this 
picture.

\subsection{B Model - Higgs Branch}

In general, the Higgs branch of an $N=4$ supersymmetric gauge theory
in three dimensions is given by a hyperk\"ahler quotient. Recall that
a hyperk\"ahler quotients are manifolds one constructs from a given 
hyperk\"ahler manifold $M$  with an action of  group $G$ that preserves the 
hyperk\"ahler structure\cite{roc1}. Associated to such a group
action are three moment maps $\mu_i:M\rightarrow g^{\ast}$, one
for each k\"ahler form, where $g^{\ast}$ is the dual of
the Lie algebra $g$ of $G$. The hyperk\"ahler quotient is defined as
the Riemannian quotient $\mu^{-1}(\vec{\zeta})/G$, where $\vec{\zeta}$
is a three vector with values in the center of  $g^{\ast}$. In 
three-dimensional $N=4$ gauge theories, one obtains a set of
equations that determine the classical vacua by integrating
out the auxiliary fields, and requiring the resulting potential to
vanish. If we are interested in the Higgs branch we put the
vevs of the scalars in the vector multiplet equal to zero,
in the case of mixed branches we can take them equal to
some other fixed value.
In this case, we obtain a real equation from the D-terms in the lagrangian, 
and a complex equation from the F-terms.
These together constitute the three equations $\vec{\mu}(x)=\vec{\zeta}$,
that also appear in the hyperk\"ahler quotient. 
The manifold $M$ is spanned by the 
vector space of
scalars in the hypermultiplets,
which is hyperk\"ahler in view of the $N=4$ supersymmetry.
The components
of $\vec{\zeta}$ correspond to the Fayet-Iliopoulos parameters in
the lagrangian. Finally, one has to divide by the action of the
gauge group, to identify equivalent vacua, and one ends up
with a Higgs branch  which is precisely the hyperk\"ahler
quotient $\mu^{-1}(\vec{\zeta})/G$. 

In the case at hand, the equations that govern the 
Higgs branch of the B-model are the same ones that
appear in the hyperk\"ahler quotient construction
of the corresponding quiver variety \cite{kn} and read
\bea 
B_{01} B_{01}^{\dagger} -B_{10}^{\dagger} B_{10} + B_{0(n-1)} B_{0(n-1)}^{\dagger}
- B_{(n-1)0}^{\dagger} B_{(n-1)0} + Q_0 Q_0^{\dagger} - \tilde{Q}_0^{\dagger}
 \tilde{Q}_0 & = & 2 \zeta^{{\bf R}}_0 {\bf 1} \non\\
B_{12} B_{12}^{\dagger} -B_{21}^{\dagger} B_{21} + B_{10} B_{10}^{\dagger}
- B_{01}^{\dagger} B_{01}  & = & 2 \zeta^{{\bf R}}_1 {\bf 1} \non\\
& \vdots & \non\\
B_{(n-1)0} B_{(n-1)0}^{\dagger} -B_{0(n-1)}^{\dagger} B_{0(n-1)} +
 B_{(n-1)(n-2)} B_{(n-1)(n-2)}^{\dagger}  & & \label {BR}\\ 
- B_{(n-2)(n-1)}^{\dagger} B_{(n-2)(n-1)}  & = & 
2 \zeta^{{\bf R}}_{n-1} {\bf 1} \non\\
B_{01} B_{10} - B_{0(n-1)} B_{(n-1)0} +Q_0 \tilde{Q}_0
& = & \xi^{{\bf C}}_0 {\bf 1}\non\\
B_{12} B_{21} - B_{10} B_{01}  & = & \zeta^{{\bf C}}_1 
{\bf 1}\non\\
& \vdots & \non\\
B_{(n-1)0} B_{0(n-1)} - B_{(n-1)(n-2)} B_{(n-2)(n-1)} 
 & = & \zeta^{{\bf C}}_{n-1} {\bf 1} 
\label{BC}
\eea
where $B_{ij}$ is a complex matrix of size $k \times k$, 
$Q_0$ and $\tilde{Q}_0$ are
respectively a column and a row vector with $k$ entries,
 and $\zeta^{{\bf R}}_i$
and $\zeta^{{\bf C}}_i$ are real and complex parameters 
which constitute
the FI parameter associated to the $i$th diagonal
$U(1)\subset U(k)$. The vector space $V$ spanned
by the components of $B_{ij}$ and $Q_0,\tilde{Q}_0$ carries the standard metric
\be \label{eq2} 
ds^2 = \sum_{i=0}^{n-1} {\rm Tr} ( dB_{i(i+1)} dB_{i(i+1)}^{\dagger}) 
+ dQ_0^{\dagger} dQ_0 + 
 d\tilde{Q}_0 d\tilde{Q}_0^{\dagger} .
\ee
The gauge group $G=U(k)^n$ acts on $V$ and on the space ${\cal M}'$
of solutions of (\ref{BR}) and (\ref{BC}), and  the Higgs branch
is the hyperk\"ahler quotient of $V$ with respect to $G$. 

We will consider the case $\sum \zeta^{{\bf R}}_i=\sum \zeta^{{\bf C}}_i = 0$.
In  this case the hyperk\"ahler quotient is the symmetric product of $k$
ALE spaces of $A_{n-1}$ type \cite{kn}. This implies that the
manifold ${\cal M}'$ is a submanifold of the set of $G
$-orbits that intersect the 
vector space $V'\subset V$, where $V'$ is constructed by 
taking all $B_{ij}$ diagonal and 
$Q_0=\tilde{Q}_0=0$. It is easy to see that ${\cal M}'/G$ is 
the same as $({\cal M}' \cap V')/G'$, 
where $G'$ is the subgroup of $G$ that maps $V'$ onto itself. 
The subgroup $G'$ is given by
the semidirect product of $U(1)^{k(n-1)}$ and the symmetric
 group $S_k$. The latter
group acts by permuting the diagonals of all $B_{ij}$ 
simultaneously. The equations
(\ref{BR})  and (\ref{BC}) consist of $k$ copies of the 
same set of equations, and are also permuted
by $S_k$. Thus
 the Higgs branch is indeed given by the symmetric product of
$k$ copies of one and the same space. This space is determined by taking the
$B_{ij}$ in (\ref{BR}) and (\ref{BC}) to be equal to a 
complex number $b_{ij}$, and $Q_0=\tilde{Q}_0=0$, and
to divide by the group $U(1)^{n-1}$. 
The equations (\ref{BR}), (\ref{BC})
 reduce to the hyperk\"ahler quotient description
of a single ALE space of type $A_{n-1}$, as given in \cite{k}, thus confirming
that the Higgs branch is the symmetric product of $k$ ALE spaces.

It remains to compute the metric on a single ALE space. For this it is convenient
to replace each 
set of complex numbers $b_{i(i+1)},b_{(i+1)i}$ by a three vector $\vr_{i}$ and an
angular variable $\sigma_i$, $0\leq \phi_i < 2\pi$ \cite{G}. 
This change of variables is
defined as follows: Given two complex numbers $a,b$, we can introduce the
quaternion $q=a-bj$. Any quaternion can be written as $q=ce^{i\sigma}$, where
$c$ is a purely imaginary quaternion, $\bar{c}=-c$. The combination $qi\bar{q}$
does not depend on $\sigma$ and is also purely imaginary, and we can define
a vector $\vr$ by \cite{G}
\beq
\frac{1}{2} (qi\bar{q}) = r^x i + (r^y + i r^z) k
\stop
\eeq
 The flat
metric $ds^2 = da d\bar{a} +db d\bar{b}$ becomes in terms of $\sigma$ and $\vr$
\be \label{eq3}
ds^2 = \frac{1}{r} d\vr^2 + r(d\sigma^2 + \vec{\omega}\cdot d\vr)^2
\ee
where $\vec{\omega}$ has the form of a one-monopole gauge field 
and satisfies
$\vec{\nabla} \times \vec{\omega} = \vec{\nabla} (\frac{1}{r})$, see
(\ref{kahler}).

The advantage of using variables $\vr_i,\phi_i$ instead of $b_{ij}$
 is that they
linearize the moment map equations (\ref{BR}) and (\ref{BC}), 
and that the metrics in these variables are similar to the ones we 
found from the one-loop computation (\ref{jan1}). If we 
introduce a three-vector
$\vec{\zeta}_{i}\equiv(\zeta^{{\bf R}}_i,{\rm Re}(\zeta^{{\bf C}}_i),
 {\rm Im}(\zeta^{{\bf C}}_i))$, then
the moment map equations simply become
\be \label{eq4}
\vr_{i}-\vr_{i-1} = \vec{\zeta}_{i}
\ee
Thus, we can solve for all $\vr_i$ in terms of $\vr_0$,
\be \label{eq5}
\vr_i= \vr_0 + \sum_{l=1}^{i} \vec{\zeta}_l.
\ee
The general solution to (\ref{BR}), (\ref{BC}) is thus parametrized by $\vr_0$ and the 
angular variables $\sigma_i$. The metric on the
 manifold of solutions is given by
\be \label{eq6} 
ds^2 = \frac{1}{r_0} d\vr_0^2 + 
 r_0 (d\sigma_0^2 + \vec{\omega}_0\cdot d\vr_0)^2 + \sum_{l=1}^{n-1} r_l
  d\sigma_l^2
\ee
where the $r_i$ are expressed in terms of $\vr_0$ by means of (\ref{eq5}).
We next take the Riemannian quotient with respect to the group
action of $U(1)^{n-1}$, which acts on the manifold of 
solutions by means of the
vector fields $V_i=\frac{\partial}{\partial \sigma_i}
 -\frac{\partial}{\partial \sigma_{i+1}}$,
$i=0\ldots n-2$. The $U(1)^{n-1}$ symmetry can be used 
to put $\sigma_1=\ldots \sigma_{n-1}=0$,
leaving the four real coordinates $\vec{r}_0$ and
 $\sigma_0$. The vector field 
$\frac{\partial}{\partial\sigma_0}$ generates an 
isometry of (\ref{eq6}) that
commutes with the group action, and therefore also 
an isometry of the quotient.
Any four dimensional hyperk\"ahler
manifold with a $U(1)$ isometry has a metric of the form \cite{roc1,roc2,PP}
\be \label{eq7}
ds^2 = g^{-2}(\vec{r}_0) d\vec{r}_0^2 + g^2(\vec{r}_0) (d\sigma_0 + 
\vec{\omega}(\vec{r}_0)\cdot 
d\vec{r}_0 )^2
\ee
where $\vec{\omega}$ is given in terms of $g^{-2}$ by the equation
$\vec{\nabla} \times \vec{\omega} = \vec{\nabla} (g^{-2}(\vec{r}_0))$, see
(\ref{jan1}) and (\ref{kahler}).  This means
that we know the full metric once we know the
 inner product of the vector
field $V=\frac{\partial}{\partial \sigma_0}$ with
 itself. This cannot be simply
read off from (\ref{eq6}), as we still have to 
take a quotient with respect 
to $U(1)^{n-1}$. If we denote by $(,)$ the metric
 (\ref{eq6}) on the solution
space and by $(,)_H$ the metric on the quotient, 
then \`a la Dirac the following
relation holds
\be \label{eq8}
(V,V)_H = (V,V) - (V,V_i) (M^{-1})^{ij} (V_j,V)
\ee
where $M_{ij}= (V_i,V_j)$. The nonzero matrix elements of $M$ are
$M_{ii}=r_i+r_{i+1}$, $M_{ii+1}=-r_{i+1}$ and $M_{i+1i}=-r_{i+1}$. 
The determinant
of $M$ satisfies the recursion relation $M^{(n)}=(r_n+r_{n-1})M^{(n-1)} -
r_{n-1}^2 M^{(n-2)}$ which is solved by $M^{(n-1)}=\prod_{i=0}^{n-1} r_i 
\left( \sum_{i=0}^{n-1} \frac{1}{r_i} \right)$. Using this result we obtain
that the only non-vanishing matrix element of $M^{-1}$ appearing in
(\ref{eq8}) is 
\be \label{eq9}
(M^{-1})^{00} = \frac{\prod_{i=1}^{n-1} r_i 
(\sum_{i=1}^{n-1} \frac{1}{r_i} ) }{
\prod_{i=0}^{n-1} r_i (\sum_{i=0}^{n-1} \frac{1}{r_i} )}=\frac{1}{r_0} - 
\frac{1}{r_0^2 \sum_{i=0}^{n-1} \frac{1}{r_i} }
\ee
Putting everything together we obtain
\be
 g^2(\vec{r}_0) = r_0 - r_0^2 (M^{-1})^{00} = 
 \frac{1}{\sum_{i=0}^{n-1} \frac{1}{r_i} }
\ee
and finally
\be
g^{-2}(\vec{r}_0) = \sum_{i=0}^{n-1} \frac{1}{r_i}.
\ee 
Using (\ref{eq5}) and comparing with the one-loop result 
(\ref{one}) with $m_{adj}=0$ we have 
\be
g^{-2}(\vec{r}_0)
= \sum_{i=0}^{n-1} 
\frac{1}{|\vec{r}_0- \vec{m}_i|}
\stop
\ee
where $\vec{m}_i=\sum_{l=1}^i \vec{\zeta}_l$. Up to a constant
shift with $\vec{\zeta}_0$ this is precisely the mirror map 
(\ref{map}). The fact that the one-loop metric on the Coulomb branch
is positive definite and smooth for generic masses strongly
suggests there are no monopole corrections to the metric
on the Coulomb branch, and that the one-loop result is
exact. In that case, both the exact Coulomb branch of the A model 
(in the infrared) and the exact Higgs branch of
B model are given by a symmetric product of ALE spaces of 
type $A_{n-1}$, 
and the relation between the masses of the hypermultiplets
in the fundamental representation in A model  and the FI
parameters in the B model is given by (\ref{map}) with $\vm_{adj}=0$.


\subsection{The Mirror Map}

The above derivation of the mirror map was restricted 
to the case when the adjoint mass in the 
A-model and the sum of the FI terms in the B-model
 were set to zero.
Consider now the case where the adjoint mass is 
different than zero. The mirror map for the adjoint mass
can be generally written as 
\beq
F(\vm_{adj},\vm_{fund}) =  \sum_{l=0}^{n-1} \vec{\zeta}_l
\stop
\eeq
If we assume that $F$ is analytic at $\vm_{adj}=\vm_{fund}=0$, then
dimensional analysis, the requirement for the correct transformation under 
the global symmetry $SU(2)_L
\times SU(2)_R$,
 and the requirement for a finite limit as 
$\vm_{fund}\rightarrow 0$ force $F$ to be linear.
We also know that $F(0,\vm_{fund})=0$, and this 
implies that $F$ is proportional to $\vm_{adj}$,
in agreement with (\ref{map}).
In principle there is also  a possibility that the
 mass of the adjoint will modify the mirror map for
the fundamental hypermultiplets. This possibility 
will be excluded in the next section by a detailed study
of the correspondence between the mass parameters 
of the A-model and the FI parameters of the B-model,
and this will also fix the relative normalization of $F$
with respect to the fundamental masses. 
 
The fact that 
the relation between the mass and the FI parameters is 
linear is also expected by the 
following reasoning: The FI parameters $\vec{\zeta}_l$
 of the B-model are 
given by the periods of the three
covariantly constant two-forms $\vec{\omega}$
 of the Higgs branch \cite{dh}
\beq
\vec{\zeta}_l = \int_{\Sigma_l}\vec{\omega}
\comma
\eeq
where  $\{\Sigma_l\}$ is a basis for the second 
homology group of the Higgs branch.
By duality it is the
Coulomb branch of the A-model.
It was argued in \cite{sw} \footnote{The argument given
in  \cite{sw} was for the $SU(2)$ gauge 
group but it can be generalized 
at least to some of the higher rank groups
such as $Sp(k)$. In fact our derivation of the mirror map shows 
that it is correct for $U(k)$ gauge groups too.}
that the periods are linear in the masses and thus we expect
a linear relation between the mass
 parameters of the A-model and the FI parameters
of the B-model and vice versa.

Finally, we note that there exists another viewpoint on 
the mirror map for the mass parameters
of the hypermultiplets in the fundamental representation
which will prove to be useful for other gauge groups.
According to theorem~2.8 in \cite{Nakajima2}, 
if $\sum \vec{\zeta}_l=0$, the Higgs branch
of the B-model develops a singularity 
if $\vec{\zeta}_k+\ldots+\vec{\zeta}_l=0$, for
$1 \leq k \leq l \leq n-1$, corresponding to
the positive roots of $A_{n-1}$
(The general case is given in (\ref{genericB}).)
On the other hand, by inspection of the one-loop
metric (\ref{jan1}) with $\vm_{adj}=0$ we see
that we expect a singularity whenever $\vm_i-\vm_j=0$.
In order for these singularities to be in
one-to-one correspondence with the singularities
in the Higgs branch of the B-model, we need
(up to an overall factor) the relations
\beq
\vec{m}_{i} - \vec{m}_{i-1}=  \vec{\zeta}_{i},~~~~~~~~~~i=1,...,n-1.
\label{mzeta}
\eeq
Equation (\ref{mzeta}) is equivalent to the mirror map (\ref{map})
with $\vm_{adj}=\sum \vec{\zeta}=0$.

\section{Duality for $U(k)$ Gauge Theories II:
  Structure of The Moduli Space of Vacua}

In this section, we analyze the moduli spaces of vacua
for various choices of mass and Fayet-Iliopoulos terms.
In general, if mass terms are turned on,
some of the Higgs branches are reduced. Conversely,
some of the Coulomb branches are reduced
by Fayet-Iliopoulos terms which, by turning on Higgs vevs,
break part of the gauge symmetry.
Here, we consider the case where
 we turn on masses of the A model and 
Fayet-Iliopoulos
terms of the B model.
We will observe a complete agreement between  the 
dimensions of various
Higgs branches of the A model and various 
Coulomb branches of the B model, provided that the 
masses and
Fayet-Iliopoulos terms are related via the mirror map 
(\ref{map}).
This result provides strong evidence for the proposed 
duality and excludes possible corrections to the
mirror map. 
Use of the proposed duality, in turn, makes it possible to 
determine
how various branches touch each other.

\subsection{ Classical Moduli Space of Vacua of The A 
Model}

In this subsection, we classify moduli spaces of
hypermultiplet using classical
arguments.
Although there are possible quantum corrections
to the way they intersect the moduli space of vector
multiplet, the metric on them will not be corrected.
Also, the structure of mixed branches will get corrected
in the direction of vector multiplet but their dimensions
will not, and we will count them.

The moduli space of hypermultiplet with its metric
is obtained by a hyperk\"ahler quotient based on
classical data.
Let $A_1=({A_1}^a_{\,\,b})$, $A_2=({A_2}^a_{\,\,b})$;
$1\leq a,b\leq k$ be a hypermultiplet in the adjoint 
representation
of $U(k)$
and $Q=(Q^a_{\,i})$, $\tilQ=(\tilQ^j_{\,b})$;
$1\leq a,b\leq k$, $0\leq i,j\leq n-1$ be 
$n$-hypermultiplets in
the fundamental representation of $U(k)$.
($Q$ and $\tilQ$ transform under $U(k)\times SU(n)$
as $({\bf k},{\bf n^*})$ and $({\bf k^*},{\bf n})$
respectively.)
The classical equations determining the vacua are
\beqa
&&[A_1,A_1^{\dag}]+[A_2,A_2^{\dag}]
+QQ^{\dag}-\tilQ^{\dag}\tilQ=0,\label{ADHMR}\\[0.1cm]
&&[A_1,A_2]+Q\tilQ=0,\label{ADHMC}\\[0.2cm]
&&\vec{\phi}Q-Q\vec{m}=0,\qquad 
\tilQ\vec{\phi}-\vec{m}\tilQ=0,
\label{MQ}\\[0.1cm]
&&[\vec{\phi},A_1]-\vm_{adj} A_1=0,\quad 
[\vec{\phi},A_2]+\vm_{adj} A_2=0
\label{MA}
\eeqa
In the above expressions, 
$\vec{\phi}=(\phi^1,\phi^2,\phi^3)$
denotes the scalers of the $U(k)$ vector multiplet
and $\vec{m}=(m^1,m^2,m^3)$ is the mass matrix.
By N=4 supersymmetry, they can be diagonalized
\beq
\vec{\phi}
=\left(
\begin{array}{ccc}
\vec{r}_1\!\!&& \\[-0.1cm]
&\!\ddots\!& \\[-0.1cm]
&&\!\!\vec{r}_k
\end{array}
\right),\qquad
\vec{m}
=\left(
\begin{array}{ccc}
\vec{m}_0&& \\
&\ddots& \\
&&\vec{m}_{n-1}
\end{array}
\right).
\eeq
Note that the trace part of $\vec{m}$ can be absorbed by a 
shift of
$\vec{\phi}$.
As we discussed before, the structure of vacua is 
substantially influenced by
the bare mass $\vm_{adj}$ of the adjoint hypermultiplet.
When $\vm_{adj} =0$,
the diagonal elements of $A_1$, $A_2$
are always massless, while there is no such flat 
direction if $\vm_{adj}\ne 0$.
We will treat the cases $\vm_{adj}=0$ and $\vm_{adj} \ne 0$ 
separately.

\subsubsection{Vanishing Adjoint Mass: $\vm_{adj}=0$}

\bigskip
As a warm-up example,
we start with the case of $n=1$.
In theories with a single flavor,
the fundamental hypermultiplet cannot have non-zero vev,
$Q=\tilQ=0$, which follows from the equations
(\ref{ADHMR}) and (\ref{ADHMC}).
The equations also imply that $A_1$ and $A_2$ can be 
diagonalized
simultaneously with $\vec{\phi}$.
Let $(\C_H^2)^k$ denote the set of eigenvalues of $A_1$ 
and $A_2$:
\beq
\left(
\left(
{z^{(1)}_1\atop z^{(1)}_2}
\right),
\ldots ,
\left(
{z^{(k)}_1\atop z^{(k)}_2}
\right)
\right)\,
\longleftrightarrow
\,
A_1=\left(
\begin{array}{ccc}
z^{(1)}_1\!\!&& \\[-0.1cm]
&\!\ddots\!& \\[-0.1cm]
&&\!\!z^{(k)}_1
\end{array}
\right),\,
A_2=\left(
\begin{array}{ccc}
z^{(1)}_2\!\!&& \\[-0.1cm]
&\!\ddots\!& \\[-0.1cm]
&&\!\!z^{(k)}_2
\end{array}
\right).
\eeq
If $\vec{\phi}$ is generic, the moduli space of 
hypermultiplet
is given by
\beq
{\cal M_H}= (\C_H^2)^k.
\label{genericMH}
\eeq
If $\vec{\phi}$ is invariant under some subgroup, say 
$G$,
of the Weyl group which acts by permuting the diagonal 
entries, then
the moduli space of hypermultiplet is $(\C_H^2)^k/G$.
Massless photons live in the subgroup of the gauge group
which is unbroken by the vevs of scalar fields.
Since the $U(1)^k$ subgroup is unbroken in the present 
case,
there are $k$-flat directions for the vector multiplets.

From here on, we will consider the case with $n\geq 2$.
Equations
(\ref{ADHMR}) and (\ref{ADHMC}) are the same as the ADHM 
equations
for the 
construction of $SU(n)$ instantons on $\R^4$ of 
instanton number $k$
\cite{Donaldson}.
Thus, if the mass constraints (\ref{MQ}) and (\ref{MA}) 
were absent,
the moduli space of hypermultiplet would be the moduli 
space of $k$-$SU(n)$ instantons on $\R^4$.
More precisely,
a solution of (\ref{ADHMR}) and (\ref{ADHMC})
describes genuine $k$-instantons only if
a condition on the rank of the matrices
$Q$, $\tilQ$, $A_1$ and $A_2$
is satisfied\footnote{
The condition is:
for any $\lambda,\mu\in\C$, both
$(A_1+\lambda,A_2+\mu,{}^t\tilQ)$
and $(\lambda-A_1,A_2-\mu,Q)$ have maximal rank $k$ (See
\cite{Donaldson}).}.
However, we take into account
all possible vacua including those
which do not meet such a condition.
A degenerate solution describes a configuration containing
a number of small instantons, the so-called ideal instantons
(see section~3.4 of \cite{dkr}). Thus, the moduli space
of hypermultiplet is in fact the moduli space
$\bM_k(SU(n))$ of ideal instantons of instanton number $k$.
This includes as subspaces the moduli spaces 
${\cal M}_{k-\ell}(SU(n)) \times Sym^{\ell} {\bf R}^4$
where $\ell$ of the instantons are small. Their
positions are labeled by
${\bf R}^4$.
If we turn on $\vec{\phi}$ and the masses $\vec{m}$
(and also $\vm_{adj}$),
the mass constraints  (\ref{MQ}) and (\ref{MA}) reduce 
the moduli space of
hypermultiplets to (a finite cover of)
a certain subspace of $\bM_k(SU(n))$.

For generic values of $\vec{\phi}$,
the gauge group $U(k)$ is broken to $U(1)^k$, and
quarks and off-diagonal part of adjoint hypermultiplet 
acquire mass.
Therefore the flat direction is $Q=\tilQ=0$, $A_1=A_2=$ 
diagonal,
and the moduli space of hypermultiplet is
given by $(\C_H^2)^k$. As the gauge symmetry $U(1)^k$
is unbroken on such vacua, we have a mixed branch
with $d_H=k$ and $d_V=k$ flat directions of hyper and 
vector multiplets.

\bigskip
\noindent
{\bf Vanishing Quark Mass}\\
We will consider first the case $\vec{m}=0$ where the theory 
possesses
global $SU(n)$ symmetry.

\medskip
At the special point $\vec{\phi}=0$, the mass constraint 
is trivial 
and
the moduli space of hypermultiplets is the full moduli 
space of ideal instantons
\beq
{\cal M_H}=\bM_k(SU(n)).
\label{spMH4A}
\eeq
This has (quaternionic) dimension $nk$. The global 
$SU(n)$ symmetry
is generically spontaneously broken but
remains unbroken on the locus $Sym^k(\C_H^2)\subset 
\bM_k(SU(n))$
of vanishing squark vevs $Q=\tilQ=0$.
The gauge group $U(k)$ is generically completely broken, 
and thus,
the moduli space (\ref{spMH4A}) is an isolated Higgs 
branch.

\medskip
Let us consider a more general value
\beq
\vec{\phi}={\rm 
diag}(0,\ldots,0,\vec{r}_{k_0+1},\ldots,
\vec{r}_k).
\eeq
If the non-zero entries are generic,
the gauge symmetry is broken to
$U(k_0)\times U(1)^{k-k_0}$, and
$A_1,A_2$ and $Q,\tilQ$ are constrained to be a 
$U(k_0)\times U(1)^{k-k_0}$
adjoint and $U(k_0)$ fundamental hypermultiplets with 
$n$ flavors respectively.
Thus, the moduli space of hypermultiplets is
\beq
{\cal M_H}=\bM_{k_0}(SU(n))\times (\C^2_H)^{k-k_0},
\label{sp2MH4A}
\eeq
which has dimension $d_H=nk_0+k-k_0$. At generic point 
on this space,
the gauge group $U(k_0)\times U(1)^{k-k_0}$ is broken to
$U(1)^{k-k_0}$. Thus, the moduli space (\ref{sp2MH4A})
extends to a mixed branch
in the $d_V=k-k_0$ flat directions for the vector multiplets.
At values of $\vec{\phi}$
whose non-zero entries are invariant under a group $G$ 
of permutations,
the factor $(\C^2_H)^{k-k_0}$ is replaced by the quotient
$(\C^2_H)^{k-k_0}/G$.

\medskip
To summarize, we list the dimensions $d_H$ and $d_V$
of the mixed branches:
\begin{center}
\renewcommand{\arraystretch}{1.5}
\begin{tabular}{|c@{\quad\vrule 
width0.8pt\quad}c|c|c|c|c|}
\hline
$\,\,d_H$&$\!\! k$&$n+k-1$&$\cdots$&$nk-n+1$&$nk$\\
\noalign{\hrule height 0.8pt}
$\,\,d_V$&$\!\! k$&$k-1$&$\cdots$&$1$&$0$\\
\hline
\end{tabular}\\[0.5cm]
Table 7: Mixed branches for $\vm_{adj}=0, \vec{m}_{fund}=0$
\end{center}
\noindent
{\bf Non-Vanishing Quark Mass}\\
We consider the case
\beq
\vec{m}={\rm 
diag}(\underbrace{\vec{m}_1,\ldots,\vec{m}_1}_{n_1},
\ldots,
\underbrace{\vec{m}_s,\ldots,\vec{m}_s}_{n_s}), \quad 
\vec{m}_i\ne\vec{m}_j
\quad i\ne j
\label{m1}
\eeq
in which the global symmetry $SU(n)$ is broken to
$SU(n_1)\times\cdots\times SU(n_s)$.
We assume here $n_i\geq 2$ but other cases can also be 
worked out.

The most general choice of $\vec{\phi}$ is
\beq
\vec{\phi}={\rm diag}(\overbrace{\vec{m}_1,\ldots,
\vec{m}_1}^{k_1},
\ldots,\overbrace{\vec{m}_s,\ldots,\vec{m}_s}^{k_s},
\vec{r}_{k_1+\cdots +k_s+1},
\ldots,\vec{r}_k)
\stop
\eeq
If $\vec{r}_{k_1+\cdots+k_s+1},\ldots,\vec{r}_k$ are 
generic,
the gauge group $U(k)$ is broken to the subgroup
$U(k_1)\times \cdots\times U(k_s)\times 
U(1)^{k-k_1-\cdots -k_s}$.
Due to  equations (\ref{MQ}) and (\ref{MA}),
the hypermultiplets $A_1,A_2$ and $Q,\tilQ$ are 
constrained respectively
to be the hypermultiplet in the adjoint representation 
of this subgroup
and $U(k_i)$ fundamental
hypermultiplets with flavors $n_i$, $i=1,\ldots, s$.
The moduli space of hypermultiplets at this point is 
thus
\beq
{\cal M_H}=\bM_{k_1}(SU_{n_1})\times\cdots\times\bM_{k_s}(SU_{n_s})
\times
(\C_H^2)^{k-k_1-\cdots -k_s},
\label{spMH4Am}
\eeq
which has dimension $d_H=\sum n_ik_i+k-\sum k_i$.
Generically on this moduli space, the gauge group is 
broken to
$U(1)^{k-k_1-\cdots -k_s}$.
Thus, the moduli space (\ref{spMH4Am})
extends to a mixed branch which has dimensions
\beq
\begin{array}{cl}
&d_H\,=\,n_1k_1+\cdots +n_sk_s+k-k_1-\cdots 
-k_s\\[0.2cm]
&d_V\,=\,k-k_1-\cdots -k_s
\end{array}
\label{DimA}
\eeq
in the directions of hyper and vector multiplets 
respectively.

\subsubsection{Non-Vanishing Adjoint Mass: $\vm_{adj}\ne 0$}

Consider next the case where $\vm_{adj}\ne 0$.
We start again with the single flavor case $n=1$.
It follows from the ADHM equations (\ref{ADHMR}) and 
(\ref{ADHMC})
that $Q=\tilQ=0$ and that $A_1$ and $A_2$ are 
diagonalizable.
On the other hand, the mass constraint
(\ref{MA}) shows that
$A_1$ and $A_2$ are nilpotent for any choice 
of $\vec{\phi}$;
we conclude that $A_1=A_2=0$.
Thus, hypermultiplets do not have a flat direction
for any value of $\vec{\phi}$ and there is only a Coulomb 
branch of dimension
$k$.

\medskip
For $n\geq 2$ one can also turn on the quark mass 
$\vec{m}$. However, we will mainly treat the case with $\vec{m}=0$
where the theory has global $SU(n)$ symmetry.
Later we make a few comments on the case $\vec{m}\ne 0$.

\bigskip
\noindent
{\it Coulomb Branch}

\medskip
For generic values of $\vec{\phi}$, quarks get mass and 
decouple $Q=\tilQ=0$.
We can also show $A_1=A_2=0$ by repeating the above 
argument.
Thus, we see that
there is no flat direction for the hypermultiplets.
Since $U(1)^k$ is unbroken, we have a Coulomb branch of 
dimension
$k$.

\bigskip
\noindent
{\it Higgs and Mixed Branches with $A_1=A_2=0$}

\medskip
At the special point $\vec{\phi}=0$,
massive adjoint hypermultiplet decouples $A_1=A_2=0$
but the quarks do not.
This is the case of QCD with $U(k)$ gauge group and $n$ flavors\footnote{
The moduli space of hypermultiplet of
$N=2$ $SU(N_c)$ QCD in four dimension was analyzed in 
\cite{APS}.}.
Using the $U(k)\times SU(n)$ rotations,
a solution to the vacuum equations 
$QQ^{\dag}-\tilQ^{\dag}\tilQ=0$
and $Q\tilQ=0$ can be expressed as
\beq
Q=\left(
\begin{array}{cccccccc}
q_1\!\!&&&\!0\!\!&&&\!0\!\!&\\[-0.2cm]
&\!\ddots\!&&&\!\ddots\!&&&\!\ddots\\[-0.2cm]
&&\!\!q_r\!&&&\!\!0\!&&\\[-0.2cm]
&&&&&&&
\end{array}
\right),\quad
{}^t\tilQ=\left(
\begin{array}{cccccccc}
0\!\!&&&\!q_1\!\!&&&\!0\!\!&\\[-0.2cm]
&\!\ddots\!&&&\!\ddots\!&&&\!\ddots\\[-0.2cm]
&&\!\!0\!&&&\!\!q_r\!&&\\[-0.2cm]
&&&&&&&
\end{array}
\right)
\label{solQ}
\eeq
for some $r$, where $q_1,\ldots, q_r$ are real 
non-negative numbers.
Note that the maximum number that $r$ can take is $k$ if 
$n\geq 2k$
and $[{n\over 2}]$ if $n< 2k$.
Let $\HH_r$ be the moduli space of hypermultiplets
consisting of vacua with rank $\leq r$ squark vevs.
The global symmetry $SU(n)$ is broken to $SU(n-2r)\times 
U(1)^r$
and there are $4nr-4r^2-r$ Nambu-Goldstone bosons.
Since there are $r$ real parameters, the moduli space 
$\HH_r$
has (quaternionic) dimension $nr-r^2$.
Remark that
$\HH_r$ is obtained by hyperk\"ahler quotient of a $nr$ 
dimensional vector space
by the completely broken subgroup $U(r)$,
and the dimension is given by the naive counting:
$\dim \HH_r=nr-\dim U(r)$.
This turns out to be a useful method to count
the dimension in complicated situations which we will
encounter in the following.
Since the gauge group is broken to $U(k-r)$, $\HH_r$ 
extends to a mixed
branch in the $k-r$ flat directions of vector multiplet.
An isolated Higgs branch $\HH_k$ exists only
when the flavor $n$ is not less than $2k$.

\bigskip
\noindent
{\it Higgs and Mixed Branches with $A_1\ne 0$, $A_2\ne 
0$}

\medskip
We can find other type of Higgs or mixed branches at 
some values of
$\vec{\phi}$. For example let us consider
\beq
\vec{\phi}={\rm diag}(\overbrace{0,\ldots,0}^{\ell_0},
\overbrace{\vm_{adj},\ldots,\vm_{adj}}^{\ell_1}).
\eeq
At this point, gauge group is broken to $U(\ell_0)\times 
U(\ell_1)$
and some of the adjoint and fundamental hypermultiplets
remain massless. 
The mass constraints (\ref{MQ}) and (\ref{MA}) impose
the vevs to be of the following form
\beq
A_1=\left(
\begin{array}{c|c}
0&0\\ \cline{1-2}
\,\alpha\,
&0
\end{array}
\right),
\quad
A_2=\left(
\begin{array}{c|c}
0&\tilde{\alpha}\\ \cline{1-2}
\,0\,
&0
\end{array}
\right),
\quad
Q=\left(
\begin{array}{ccc}
&q&\\ \cline{1-3}
&0&
\end{array}
\right),
\quad
{}^t\tilQ=\left(
\begin{array}{ccc}
&\!{}^t\tilde{q}\,&\\ \cline{1-3}
&0&
\end{array}
\right),
\eeq
where the $k$ columns (rows) are decomposed into blocks 
of size
$\ell_0$ and $\ell_1$.
Under the local and global symmetry
$U(\ell_0)\times U(\ell_1)\times SU(n)$, 
$\alpha$ and $\tilde{\alpha}$
transform as $({\bf \ell_0}^*,{\bf \ell_1},{\bf 1})$,
$({\bf \ell_0},{\bf \ell_1}^*,{\bf 1})$
while $q$ and $\tilde{q}$
transform as $({\bf \ell_0}, {\bf 1},{\bf n}^*)$,
$({\bf \ell_0}^*,{\bf 1},{\bf n})$ respectively.
The D-term equations
with respect to $U(\ell_0)$ and $U(\ell_1)$ gauge 
symmetry
read 
\beqa
&&
qq^{\dag}-\tilde{q}^{\dag}\tilde{q}=\alpha^{\dag}\alpha
-\tilde{\alpha}\tilde{\alpha}^{\dag},\quad
q\tilde{q}=\tilde{\alpha}\alpha,
\label{DUk0}\\
&&
\alpha\alpha^{\dag}-\tilde{\alpha}^{\dag}\tilde{\alpha}=
0,\quad
\alpha\tilde{\alpha}=0.
\label{DUk1}
\eeqa
Equations (\ref{DUk1}) admit solutions as
(\ref{solQ})
and in particular require $\alpha$ and $\tilde{\alpha}$
to be of the same rank, say $k_1$. 
If we insert such a solution,
equations (\ref{DUk0}) also 
requires $q$ and $\tilde{q}$ to be of the same rank, say 
$k_0$.
In addition to the obvious bound $k_1\leq \ell_1$ and 
$k_0\leq \ell_0$,
the ranks must satisfy the relation
\beq
k_0\geq 2k_1,\qquad n\geq 2k_0-k_1.
\label{range0}
\eeq
Let $\HH_{k_0,k_1}$ be the moduli space of such vacua 
with lower rank cases being included.
It is the hyperk\"ahler quotient of a vector space
of dimension $nk_0+k_0k_1$ by the completely broken 
subgroup $U(k_0)\times U(k_1)$. 
Thus, according to the previous remark, its dimension is
$nk_0+k_0k_1-k_0^2-k_1^2$. Generically 
on this space,
the gauge group
is broken to $U(\ell_0-k_0)\times U(\ell_1-k_1)$.
Thus $\HH_{k_0,k_1}$ extends to a mixed
branch in the $k-k_0-k_1$ flat directions of vector 
multiplet.
Note that it exists
when $k_0+k_1\leq k$, $n\geq 2k_0-k_1$ and $k_0\geq 
2k_1$
(irrespectively of $\ell_0,\ell_1$).
It is an isolated Higgs branch if $k_0+k_1=k$ which is 
possible only when $n\geq k$.

\medskip
In general, flat directions of hypermultiplet
can be found for values of 
$\vec{\phi}$ whose entries are integer multiples of 
$\vm_{adj}$.
Let us consider the case in which $k_j$ entries are 
$j\vm_{adj}$ where $j$ runs over integers from $-p\leq 0$
to $q\geq 0$. There exists a non-trivial moduli space
of hypermultiplet $\HH_{\{k_i\}}$ when the $k_j$
satisfy the following conditions
\beqa
&&\sum_{i=-p}^qk_i\leq k,\qquad 2k_0-k_{-1}-k_{1}\leq n
\label{range1}\\
&&k_{i-1}-k_i\geq k_i-k_{i+1}, \quad i\ne 0.
\label{range2}
\eeqa
It extends to a mixed branch which has dimensions
\beq
d_H=nk_0+\sum_{i=-p}^{q-1} k_ik_{i+1}-\sum_{i=-p}^q 
k_i^2,\qquad 
d_V=k-\sum_{i=-p}^q k_i.
\label{dimHA}
\eeq
Note that the condition (\ref{range2}) which is a 
generalization of (\ref{range0}) means that
the plot of $k_j$ against the horizontal $j$ axis is 
concave in the regions $j>0$ and $j<0$. This
concave property will become more important in the next 
subsection.

\medskip
\noindent{\bf Generic Quark Mass}

\medskip
When two or more quark masses are coincident, quarks 
have a flat direction. 
Otherwise, a flat direction for the hypermultiplets is possible 
only when the mass constraints (\ref{MQ}) and (\ref{MA})
allow some components of the quarks and adjoint hypermultiplet
that are charged under common subgroups to be massless.
However, this cannot happen at any value of $\vec{\phi}$
if the masses are generic in the following sense
\beq
\vm_{adj}\ne 0\quad \mbox{and}\quad \vec{m}_i-\vec{m}_j\ne 
\ell \vm_{adj}\quad\mbox{for any $0\leq j<i\leq n-1$ and 
$-k<\ell<k$}.
\label{genericA}
\eeq
Conversely, when this condition
is broken, a flat direction for the  hypermultiplets
 does exist  for some value of $\vec{\phi}$.

\subsection{ Classical Moduli Space of Vacua of The B 
Model}

We look at the moduli space of vacua of the B model in 
such a way
that various Coulomb or mixed branches are emanating 
from the underlying
moduli space ${\cal  M}_H$ of hypermultiplet. As FI terms 
are turned on,
the moduli space ${\cal  M}_H$ is deformed and the Coulomb
branches get reduced.
The dimension of the moduli space of vector multiplet 
emanating
from a point of ${\cal  M}_H$ is given by the rank of the 
unbroken gauge group.
In this subsection, we characterize and classify points 
of ${\cal  M}_H$
with respect to the unbroken gauge group.

Recall that the B model has gauge group
$U(k)^n=\prod_{i=0}^{n-1}U(k)_i$ and matter 
hypermultiplets
$B_{i(i+1)}, B_{(i+1)i}$ in the ``bifundamental'' 
representation
of $U(k)_i\times U(k)_{i+1}$
and $\Qn,\tilQn$ in the fundamental representation of 
$U(k)_0$.
($B_{i(i+1)}$ and $B_{(i+1)i}$ transform as
$({\bf k},{\bf k^*})$ and $({\bf k^*},{\bf k})$ under
$U(k)_i\times U(k)_{i+1}$ respectively.)
The moduli space ${\cal  M}_H$ of hypermultiplet is 
determined at the classical
level as the set of solutions of
the classical equations (\ref{BR}), (\ref{BC}) modulo 
the $U(k)^n$ 
gauge group action.
We note that this is the same as
the hyperk\"ahler quotient construction of Hilbert 
Scheme of points
on an ALE space by Kronheimer and Nakajima 
\cite{kn,Nakajima,Nakajima2}.
We do not impose mass constraints
like (\ref{MQ}) and (\ref{MA}) on hypermultiplets.
Instead, we use them
to force the  flat directions of the vector multiplet to lie 
in the direction of the unbroken gauge group.

As we will see, the structure of vacua is greatly 
affected
by the trace part $\sum \vec{\z}_{i}$ of the FI 
parameters
$\vec{\z}=(\vec{\z}_0,\vec{\z}_1,\ldots,\vec{\z}_{n-1})$
.

\subsubsection{Tracefree FI Parameters: 
$\sum_i\vec{\z}_i=0$}

In the case where the trace part of the FI parameters 
vanishes,
one can show that $\Qn=\tilQn=0$ and $B_{ij}$
are diagonalizable at the same time
\beq
B_{ij}=
\left(
\begin{array}{ccc}
b^{(1)}_{ij}&&\\
&\ddots&\\
&&b^{(k)}_{ij}
\end{array}
\right).
\eeq
The $l^{\rm th}$ diagonal entries 
$b_l=(b^{(l)}_{ij})$
satisfy a system of equations. Namely, that of the $k=1$ 
model:
\beqa
&&|b_{i(i-1)}|^2-|b_{(i-1)i}|^2+|b_{i(i+1)}|^2
-|b_{(i+1)i}|^2=\z^{\R}_i
\label{ALER}\\
&&b_{i(i-1)}b_{(i-1)i}-b_{i(i+1)}b_{(i+1)i}=\z_i^{\C}
\label{ALEC}
\eeqa
where $\z^{\R}=\z^1,\z^{\C}=\z^2+i\z^3$ for 
$\vec{\z}=(\z^1,\z^2,\z^3)$.
The moduli space of hypermultiplet for $k=1$ model is
the quotient by $U(1)^{n-1}$ of the set of solutions of 
these equations.
(Note that the diagonal $U(1)$ subgroup of $U(1)^n$
is always unbroken, and can be forgotten upon quotient.)
This is the same as
the Kronheimer's hyperk\"ahler quotient construction 
\cite{k} of
the ALE space $X_{\vec{\z}}$ of type $A_{n-1}$.
For $k\geq 1$,
we have $k$ copies of this space and
dividing by the residual permutation symmetry $S_k$
we obtain
\beq
{\cal M}_H=Sym^k(X_{\vec{\z}}).
\eeq
At generic points of ${\cal  M}_H$, all $b_{ij}^{(l)}$
are non-zero and the gauge group $U(k)^n$ is broken to
the diagonal subgroup
$U(1)^k$ of the maximal torus $(U(1)^k)^n$.
Thus, the vector multiplet generically has $k$-flat 
directions
and there is no pure Higgs branch.
As we will see in the following,
for a non-generic choice of the FI parameters $\vec{\z}$
there are special points in ${\cal M}_H$ at which the 
unbroken gauge group
has higher rank.
This enhancement of unbroken gauge group corresponds to
the singularity of the space $X_{\vec{\z}}$.

\noindent
{\bf Turning Off FI Parameters $\vec{\z}=0$}\\
We first consider the case with FI terms turned off.
Let us look at the moduli space ${\cal  M}_H=X_{\vec{0}}$ 
for the $k=1$ model.
It follows from the equations (\ref{ALER}) and 
(\ref{ALEC}) that
$|b_{i(i+1)}|$, $|b_{(i+1)i}|$ and $b_{i(i+1)}b_{(i+1)i}$ 
are
independent of $i$. Then, we can define $z_1$ and $z_2$
by
\beqa
&&z_1z_2=b_{i(i+1)}b_{(i+1)i}\\
&&z_1^n=b_{01}b_{12}\cdots b_{(n-1)0}\\
&&z_2^n=b_{0(n-1)}\cdots b_{21}b_{10}
\eeqa
up to $\Z_n$ ambiguity $(z_1,z_2)\sim
(\e^{2\pi i\over n}z_1,\e^{-{2\pi i\over n}}z_2)$.
Thus, we see that the $k=1$ moduli space is just the 
$\Z_n$
orbifold $\C^2/\Z_n$.
Note also that by introducing gauge invariant variables
$x=z_1z_2$, $y=z_1^n$, and $z=z_2^n$, we obtain the 
standard
relation $x^n=yz$.
The $A_{n-1}$ simple singularity at the origin
corresponds to the solution $b_{ij}\equiv 0$ on which
the gauge group $U(1)^n$ is totally unbroken.
At other points, one or both of
$b_{i(i+1)}$ and $b_{(i+1)i}$ is non-vanishing for each 
$i$
and hence the gauge group is broken to the diagonal 
$U(1)$.
Thus, we have a Coulomb branch of dimension $n$ and a 
mixed branch
with a single flat direction for each of the hyper- and 
vectormultiplets.

\medskip
For general $k$,
the moduli space of hyper multiplet
is the $k^{\,\rm th}$ symmetric product
\beq
{\cal  M}_H=Sym^k(\C^2/\Z_n).
\eeq
For each $k_0$, $0\leq k_0\leq k$,
let ${\cal N}_{k_0}\subset Sym^k(\C^2/\Z_n)$ be the 
submanifold
of dimension $k-k_0$
corresponding to the set of points in $(\C^2/\Z_n)^k$
whose $k_0$ entries are the $A_{n-1}$ singularity.
A generic point in ${\cal N}_{k_0}$ corresponds to
a vacuum with $b^{(1)}=\cdots =b^{(k_0)}=0$ on which
the gauge symmetry $U(k)^n$ is broken to
the subgroup $U(k_0)^n\times U(1)^{k-k_0}$ of rank 
$nk_0+k-k_0$.
If the non-zero entries $b^{(k_0+1)},\ldots,b^{(k)}$ are 
invariant
under a group of permutations, the factor $U(1)^{k-k_0}$
is replaced by a larger group but the rank is still 
$k-k_0$.
Thus,
along the submanifold ${\cal N}_{k_0}$ of the moduli 
space
of hypermultiplet, the vector multiplet
has $nk_0+k-k_0$ flat directions. To summarize,
we list the dimensions of mixed branches where 
$d_H$, $d_V$
denotes the number of flat directions of hyper and 
vector multiplets:
\begin{center}
\renewcommand{\arraystretch}{1.5}
\begin{tabular}{|c@{\quad\vrule 
width0.8pt\quad}c|c|c|c|c|}
\hline
$\,\,{d}_H$&$\!\! k$&$k-1$&$\cdots$&$1$&$0$\\
\noalign{\hrule height 0.8pt}
$\,\,{d}_V$&$\!\! 
k$&$n+k-1$&$\cdots$&$nk-n+1$&$nk$\\
\hline
\end{tabular}\\[0.5cm]
Table 8: Mixed branches for $\vec{\z}_0=\cdots 
=\vec{\z}_{n-1}=0$
\end{center}

\noindent
{\bf Turning On Tracefree FI Parameters}\\
We consider the case
\beq
\vec{\z}=(\underbrace{\vec{\z}_1,0,\ldots,0}_{n_1},
\underbrace{\vec{\z}_2,0,\ldots,0}_{n_2},\ldots,
\underbrace{\vec{\z}_s,0,\ldots,0}_{n_s})
\label{zeta1}
\eeq
in which
\beq
\vec{\z}_1+\cdots +\vec{\z}_s=0\quad\mbox{but}\quad
\vec{\z}_{i+1}+\cdots +\vec{\z}_j\ne 0,\quad 0< 
i<j\leq s.
\eeq
This corresponds to the choice of mass (\ref{m1})
under the mirror map (\ref{map})
where $\vec{m}_i=\vec{\z}_1+\cdots +\vec{\z}_i$.

The moduli space ${\cal  M}_H=X_{\vec{\z}}$ for the $k=1$ 
model is
an orbifold with singularities of types
$A_{n_1-1},A_{n_2-1},\ldots,A_{n_s-1}$ at $s$ distinct 
points.
This can be seen by an argument as in the $\vec{\z}=0$ 
case.
For instance,
the point with $A_{n_1-1}$ singularity corresponds to 
the vacuum
with $b_{i,i+1}=b_{i+1,i}=0$ for $i=0,1,\ldots,n_1-1$,
which is invariant under the subgroup $U(1)^{n_1}$ of 
the gauge
group $U(1)^n$.
In general, at the point with $A_{n_i-1}$ singularity
the unbroken gauge group is $U(1)^{n_i}$, while it is 
the
diagonal $U(1)$ at other points.
Thus, we have $s$-Coulomb branches of dimensions 
$n_1,\ldots, n_s$
and one mixed branch of dimensions $\tild_H=\tild_V=1$.

\medskip
For general $k$, let 
${\cal N}_{k_1,\ldots,k_s}$
be the submanifold of ${\cal  M}_H=Sym^k(X_{\vec{\z}})$
corresponding to the points in $(X_{\vec{\z}})^k$
whose $k_i$ entries are the $A_{n_i-1}$ singularity.
On this submanifold, the gauge group is broken to
$U(k_1)^{n_1-1}\times\cdots\times U(k_s)^{n_s-1}\times 
U(1)^k$.
Thus, we have a mixed branch of dimensions
\beq
\begin{array}{cl}
&d_H\,=\,k-k_1-\cdots-k_s\\[0.2cm]
&d_V\,=\,n_1k_1+\cdots +n_sk_s+k-k_1-\cdots -k_s
\end{array}
\label{DimB}
\eeq
along the submanifold 
${\cal N}_{k_1,\ldots,k_s}$.

\subsubsection{FI Parameters With Non-Vanishing Trace: 
$\sum_i\vec{\z}_i\ne 0$}

When the trace $\sum\vec{\z}_i$ of the FI parameters is 
non-vanishing, things drastically change. By summing up
the equations (\ref{BR}) and (\ref{BC}) and taking the 
trace, we obtain $||Q_0||^2-||\tilQ_0||^2=2k\sum\z_i^{\R}$
and $\tilQ_0Q_0=k\sum \z_i^{\C}$, and thus $Q_0$ or 
$\tilQ_0$ cannot be zero.
In addition, the $B_{i,j}$ cannot be simultaneously 
diagonalizable.
Namely, we have lost the structure of the symmetric product 
of the moduli space $X_{\vec{\z}}$ of the $k=1$ model.
Instead, our moduli space is the hyperk\"ahler quotient 
construction of the Hilbert scheme of
$k$-points on $X_{\vec{\z}}$:
\beq
{\cal  M}_H\,=\,Hilb^{[k]}X_{\vec{\z}}.
\eeq
When $\vec{\z}$ is generic and $X_{\vec{\z}}$ is smooth, 
it is known that $Hilb^{[k]}X_{\vec{\z}}$ is a 
resolution of 
the diagonal or quotient singularities
 of $Sym^kX_{\vec{\z}}$ and is in 
particular smooth. This means that the gauge group
$U(k)^n$ is completely broken at every point of 
${\cal  M}_H$ and there is no flat direction for the vector 
multiplets.

For some special values of $\vec{\z}$ such that 
$X_{\vec{\z}}$ is singular, $Hilb^{[k]}X_{\vec{\z}}$ 
inherits the singularity
of $X_{\vec{\z}}$. At a singular point, some subgroup of 
$U(k)^n$ remains unbroken and flat directions of
vector multiplet appear.
Here we classify such unbroken subgroups for the special 
value
\beq
\vec{\z}=(\vec{\z}_0,0,\ldots,0)
\eeq
for which the $k=1$ moduli space $X_{\vec{\z}}$ is the 
orbifold $\C^2/\Z_n$ with an $A_{n-1}$ simple singularity.
Using an $SU(2)_L$ rotation, we may put 
$\vec{\z}_0=(c,0,0)$ with $c>0$.

\bigskip
\noindent
{\it Convex Graphs and Mixed Branches}

\medskip
By looking at the first equation of (\ref{BR}), we see 
that $(B_{0,n-1},B_{0,1},Q_0)$ has rank $k$ and hence 
$U(k)_0$
is always completely broken. However, the groups 
$U(k)_i$ at other sites may contain unbroken pieces.
Let us consider configurations such that $U(k)_i$ is 
broken to $U(\ell_i)_i$ and its centralizer 
$U(k-\ell_i)_i$ is completely
broken. Such configurations exist only when the $\ell_i$ 
satisfy a certain condition. Let us consider making a 
plot
of $\ell_i$ against the horizontal $i$ axis where $i$ 
runs from $0$ to $n\equiv 0$.
As we noted in subsection 5.1.2, the plot of the rank 
$k-\ell_i$ of the completely broken gauge groups
must be concave. In other words, the plot of $\ell_i$ is 
convex. 
Thus, for each convex integral graph 
$\{\ell_i\}_{i=0}^n$ with $\ell_0=\ell_n=0$, $\ell_i\leq 
k$,
we have a submanifold of ${\cal  M}_H$
with unbroken gauge group $\prod_iU(\ell_i)$.
Its dimension is 
$d_H=k+\sum_{i=0}^{n-1}(k-\ell_i)(k-\ell_{i+1})-\sum
_{i=0}^{n-1} (k-\ell_i)^2$ and
it extends to a mixed branch with dimension 
$d_V=\sum \ell_i$ in the direction of vector 
multiplet.

This result can be rephrased in the following way.
Suppose that the steepest ascending slope of the plot of 
$\ell_i$ is $q+1$, and the steepest descending slope is 
$-p-1$.
For $-p-1\leq i\leq q+1$, let $e_i$ be the number of 
steps with slope $i$. Since the plot starts with 
$\ell_0=0$
and ends with $\ell_{n}=0$, these numbers satisfy $\sum 
ie_i=0$.
Let us introduce numbers $k_i$; $-p\leq i\leq q$ by
\beq
\begin{array}{ll}
&e_{q+1}=k_q\\
&e_q=k_{q-1}-2k_q\\
&e_{q-1}=k_{q-2}-2k_{q-1}+k_q\\
&\,\,\,\,\,\cdots\cdots\\
&e_2=k_1-2k_2+k_3\\
&e_1=k_0-2k_1+k_2
\end{array}\qquad
\begin{array}{ll}
&e_{-p-1}=k_{-p}\\
&e_{-p}=k_{-p+1}-2k_{-p}\\
&e_{-p+1}=k_{-p+2}-2k_{-p+1}+k_{-p}\\
&\,\,\,\,\,\cdots\cdots\\
&e_{-2}=k_{-1}-2k_{-2}+k_{-3}\\
&e_{-1}=k_0-2k_{-1}+k_{-2}.
\end{array}
\label{slope}
\eeq
It appears that $k_0$ has two solutions $\sum_{i>0} 
ie_i$ and $\sum_{i>0} ie_{-i}$
but they coincide due to the relation $\sum ie_i=0$.
In fact, it is the highest value of $\ell_i$.
In terms of $\{k_i\}$, the dimensions $d_H$ and 
$d_V$ of the mixed branch
can be expressed as
\beq
d_H=k-\sum k_i,\qquad
d_V=nk_0+\sum k_ik_{i+1}-\sum k_i^2.
\label{2dim}
\eeq
In particular $\sum k_i\leq k$.
Since $e_i$ are non-negative integers, $k_i$ satisfy the 
concave property
\beq
k_{i-1}-k_i\geq k_i-k_{i+1},\qquad i\ne 0.
\label{2range2}
\eeq
Since the total number of steps is $n$, we have $n\geq 
\sum_{i\ne 0}e_i$, i.e.
\beq
n\geq 2k_0-k_{-1}-k_1.
\label{2range1}
\eeq
It is easy to see that each sequence $\{k_i\}_{-p\leq 
i\leq q}$ satisfying (\ref{2range1}), (\ref{2range2}) 
and
$\sum k_i\leq k$ determines a convex graph 
$\{\ell_i\}_{i=0}^n$ having $e_i$ steps of slope $i$ 
where $e_i$ is given by
(\ref{slope}).

\bigskip
\noindent
{\it Adjacency Relations}

\medskip
Let us denote by ${\cal N}_{\{k_i\}}$ the submanifold
of $Hilb^{[k]}(\C^2/\Z_n)={\cal  M}_H$ corresponding to a 
convex graph determined by the sequence $\{k_i\}$.
As we move around the moduli space ${\cal  M}_H$, unbroken 
gauge group can suddenly be enhanced but the converse 
will never occur.
This property tells us some information on how the 
submanifolds
${\cal N}_{\{k_i\}}$ are related with each other. 
Let us consider two graphs $\{\ell_i\}$ and 
$\{\ell_i^{\prime}\}$ determined by $\{k_i\}$ and 
$\{k_i^{\prime}\}$
respectively. Then, ${\cal N}_{\{k_i^{\prime}\}}$ intersects 
with a boundary of ${\cal N}_{\{k_i\}}$ only if 
$\ell_i^{\prime}\geq \ell_i$
for any $i$. It is easy to see that the latter condition 
holds
if and only if $k_i^{\prime}\geq k_i$ for any $i$ which 
we represent by $\{k_i^{\prime}\}\geq \{k_i\}$.
Thus, we have seen that
\beq
\overline{{\cal N}_{\{k_i\}}}\,\,\,\subset 
\bigcup_{\{k_i^{\prime}\}\geq 
\{k_i\}}{\cal N}_{\{k_i^{\prime}\}}.
\eeq

\bigskip
\noindent
{\bf Generic Values of FI Parameters}

\medskip
According to \cite{Nakajima2} theorem 2.8, 
$Hilb^{[k]}X_{\vec{\z}}$ is smooth when the FI parameter 
$\vec{\z}$
satisfies a certain condition. In our language it reads 
as
\beq
\sum_h \vec{\z}_h\ne 0\quad\mbox{and}\quad
\vec{\z}_i+\cdots+\vec{\z}_j\ne \ell 
\sum_h\vec{\z}_h\quad\mbox{for any $1\leq j\leq i\leq 
n-1$, $-k<\ell<k$.}
\label{genericB}
\eeq
When this condition is satisfied, gauge group is 
completely broken everywhere and there is only a
Higgs branch of dimension $k$.

\subsection{ The Mirror Map Revisited}

In subsection 5.1, we determined and
classified the various moduli spaces of hypermultiplet
emanating from the classical moduli space of vector 
multiplet.
In subsection 5.2, we classified submanifolds of
the moduli space ${\cal  M}_H$ of hypermultiplet with 
respect to
the rank of the unbroken gauge group.
If we compare the results,
we can see an agreement of dimensions of mixed branches
\beq
(d_H,d_V)_{\rm A-model}\,=\,(d_V,d_H)_{\rm B-model}
\eeq
provided that masses and FI parameters are related
under the mirror map (\ref{map}).
For example, compare

$\bullet$ Table 7 for $\vm_{adj}=\vec{m}\equiv 0$ and Table 8 
for 
$\vec{\z}_i\equiv 0$

$\bullet$ Dimensions (\ref{DimA}) for the mass 
(\ref{m1}) and (\ref{DimB})
for the FI parameter (\ref{zeta1})

$\bullet$ Dimensions (\ref{dimHA}) with (\ref{range1}),
(\ref{range2}) for $\vm_{adj}\ne 0,\vec{m}=0$ and

\vspace{-0.2cm}
\hspace{0.2cm} Dimensions (\ref{2dim}) with 
(\ref{2range1}), (\ref{2range2})
for $\vec{\z}_0\ne 0,\vec{\z}_{i>0}=0$

$\bullet$ Condition (\ref{genericA}) for the mass to be 
generic and Condition (\ref{genericB}) for the FI

\vspace{-0.2cm}
\hspace{0.3cm} parameters to be generic.

\noindent
This agreement gives strong evidence of our duality 
proposal. In particular, the third one excludes the 
possibility of non-trivial dependence
of the trace $\sum\vec{\z}_j$ in $\vec{m}_i$ like
$\vec{m}_i=\sum_{l=0}^i 
\vec{\z}_l+c_i\sum_{j=0}^{n-1}\vec{\z}_j$.
Also, the last one shows that absence of a flat direction 
for the hypermultiplets corresponds to the smoothness of
$Hilb^{[k]}X_{\vec{\z}}$ only when the mirror map is 
normalized as in (\ref{map}).
Thus, we have excluded all possible corrections to the
mirror map (\ref{map}) and completed the
proof of it.

\subsection{ Quantum Moduli Space of Vacua}

In this subsection, which is mostly a summary of the
results we obtained so far,
we give a description what the quantum moduli space of 
vacua of the A model looks like if our duality 
conjecture is assumed to be correct.
In particular,
we locate the moduli spaces of hypermultiplet
on the {\it quantum}
moduli space of vector multiplet $\MV$
by identifying the latter with
the moduli space of hypermultiplet ${\cal M}_H$
of the B model.

\bigskip
\noindent
{\bf The Self-Dual Model}\\
When there is only a single flavor $n=1$, the A model 
coincides with
the B model and therefore is expected to be self-dual.
The model has two parameters: the bare mass $\vm_{adj}$ of 
the adjoint
hypermultiplet
and the FI parameter $\vec{\z}$ for the unique $U(1)$ 
factor of
the gauge group.

\medskip
\underline{$\vm_{adj}=0, \vec{\z}=0$}

\medskip
In this case, quantum moduli space of vector multiplet 
is
$\MV=Sym^k(\C^2_V)$ where $\C_V^2=X_{\vec{0}}$
is the quantum moduli space for the $k=1$ model.
At the generic point of $\MV$ represented by a point of 
$(\C_V^2)^k$
whose entries are distinct with each other,
we have $(\C_H^2)^k$ as the moduli space of 
hypermultiplet.
When the representative in $(\C_V^2)^k$ is
invariant under a group $G$ of permutations,
the moduli space of hypermultiplet collapses to
$(\C_H^2)^k/G$. Thus, the quantum moduli space is given 
by
\beq
\M_{\rm total}=Sym^k(\C_V^2\times \C_H^2).
\eeq

\medskip
\underline{$\vm_{adj}\ne 0, \vec{\z}=0$}

\medskip
When $\vm_{adj}\ne 0$, there is a monopole correction that 
smooths out the
singularity due to $S_k$ quotient, and the quantum 
moduli space of vector 
multiplet is the Hilbert scheme of $k$-points on 
$\C_V^2$:
$\MV=Hilb^{[k]}(\C_V^2)$.
Since $\vm_{adj}\ne 0$ meets the condition
(\ref{genericA})
to be generic,
the flat directions of hypermultiplet are completely 
lifted.

\medskip
\underline{$\vm_{adj}=0, \vec{\z}\ne 0$}

\medskip
Since $\vec{\z}\ne 0$ meets the condition
(\ref{genericB}) to be generic,
the flat directions of vector multiplet are completely 
lifted
and we have a single smooth Higgs branch
which is again the Hilbert scheme of points 
$Hilb^{[k]}(\C_H^2)$.

\medskip
In summary, we list the quantum moduli space of vacua:

\begin{center}
\renewcommand{\arraystretch}{1.5}
\begin{tabular}{|c@{\quad\vrule width0.8pt\quad}c|c|}
\hline
&Moduli Space&$(d_V,d_H)$\\
\noalign{\hrule height 0.8pt}
$\vm_{adj}=0,\,\vec{\z}=0\!\!$&$\,\,Sym^k(\C_V^2\times\C_H^2)
\,
\,$&
$(k,k)$\\
\hline
$\vm_{adj}\ne 
0,\,\vec{\z}=0\!\!$&$Hilb^{[k]}\C_V^2$&$(k,0)$\\
\hline
$\vm_{adj}=0,\,\vec{\z}\ne 
0\!\!$&$Hilb^{[k]}\C_H^2$&$(0,k)$\\
\hline
\end{tabular}\\[0.5cm]
Table 9: Quantum Moduli Space of Vacua of the $n=1$ 
Model
\end{center}

\noindent
{\bf Multi Flavor Case}

\medskip
\underline{$\vm_{adj}=0,\,\vec{m}=0$}

\medskip
When all the mass terms are turned off, the moduli space 
of vector
multiplet is given by $\MV=Sym^k(\C^2/\Z_n)$
which decomposes into $k+1$ submanifolds 
${\cal N}_{k_0}$,
$0\leq k_0\leq k$.
Recall that ${\cal N}_{k_0}$ corresponds to the set of 
points
in $(\C_V^2/\Z_n)^k$ whose $k_0$ entries are the 
$A_{n-1}$ singularity.
The moduli space of hypermultiplet emanating from
a generic point of ${\cal N}_{k_0}$ is
$(\C_H^2)^{k-k_0}\times\bM_{k_0}(SU(n))$.
At the point whose representative in $(\C^2/\Z_n)^k$
is invariant under a group $G\times S_{k_0}$ of 
permutations,
the moduli space of hypermultiplet
collapses to $(\C_H^2)^{k-k_0}/G\times\bM_{k_0}(SU(n))$.
Thus, we have located
the moduli spaces of hypermultiplet on the submanifold 
${\cal N}_{k_0}$.
The resulting mixed branch, including its boundary,
is given by
\medskip
\beq
\M_{k_0}\,=\, 
Sym^{k-k_0}(\C^2/\Z_n\times \C^2_H)
\times
\bM_{k_0}(SU(n)).
\eeq
It has dimensions $d_V=k-k_0$ and $d_H=nk_0+k-k_0$.
The quantum moduli space is
now represented as a union of these branches:
\medskip
\beq
\M_{\rm total}\,\,=\bigcup_{0\leq k_0\leq k}\M_{k_0}\,.
\eeq
Note that $\M_k=\bM_k(SU(n))$
is a unique Higgs branch of dimension $nk$. The ``basic 
branch''
\beq
\M_0\,=\,Sym^k(\C^2/\Z_n\times \C^2_H)
\eeq
is a mixed branch of dimension $2k$ which has no 
non-trivial $SU(n)$ action.
On any other branch $\M_{k_0}$,
the global $SU(n)$ symmetry is generically spontaneously 
broken
due to squark vevs.
It touches the basic branch $\M_0$
along the submanifold of dimension $2k-k_0$
of $SU(n)$-fixed points.
The theories in that submanifold have unbroken $SU(n)$ 
symmetry.
The branches $\M_{k_0}$ with $k_0\geq 1$ also touch
each other; a boundary of $\M_{k_0}$ is embedded in
$\M_{k_0+\ell}$ according to the embedding 
of
$Sym^{\ell}(\C_H^2)\times \bM_{k_0}(SU(n))$ in
$\bM_{k_0+\ell}(SU(n))$.

\bigskip
\underline{$\vm_{adj}=0,\,\vec{m}\ne 0$}

\medskip
We consider the case with the bare mass $\vec{m}$ being 
given by
(\ref{m1})
in which the theory has global symmetry
$SU(n_1)\times \cdots\times SU(n_s)$.
The moduli space of vector multiplet is 
$\MV=Sym^k(X_{\vec{\z}(m)})$
where $\vec{\z}(m)$ is mirror image of 
$\vec{m}$.

The quantum moduli space is represented as:
\beq
\M_{\rm total}\,\,=
\bigcup_{k_i\geq 0 \atop k_1+\cdots +k_s\leq 
k}\M_{k_1,\ldots,k_s}\,.
\eeq
where
\beq
\M_{k_1,\ldots,k_s}=
Sym^{k-\sum k_i}(X_{\vec{\z}(m)}\times\C_H^2)
\times\bM_{k_1}(SU_{n_1})\times\cdots\times\bM_{k_s}(SU_
{n_s})
\eeq
is a mixed branch of dimensions $d_V=k-\sum k_i$ and
$d_H=\sum n_ik_i+k-\sum k_i$.
The basic branch
\beq
\M_{0,\ldots,0}=Sym^k(X_{\vec{\z}(m)}\times\C_H^2)
\eeq
has no non-trivial $SU(n_1)\times\cdots \times SU(n_s)$ 
action.
Any other branch $\M_{k_1,\ldots,k_s}$
has non-trivial action of this group
and touches the basic branch $\M_{0,\ldots,0}$ along the 
submanifolds
of fixed points.
Theories in the fixed point submanifold
have unbroken $SU(n_1)\times\cdots \times SU(n_s)$ 
global symmetry.

\bigskip
\underline{$\vm_{adj}\ne 0,\,\vec{m}=0$}

\medskip
Finally, let us consider the case $\vm_{adj}\ne 0$ and 
$\vec{m}=0$ in which the theory possesses global $SU(n)$ 
symmetry.
This choice of mass is mapped to the FI parameter 
$\vec{\z}=(\vm_{adj},0,\ldots,0)$.
The moduli space of vector multiplet is 
$\MV=Hilb^{[k]}(\C^2/\Z_n)$,
the Hilbert scheme of $k$-points on $\C^2/\Z_n$.

\medskip
The quantum moduli space is represented as follows:
\beq
\M_{\rm total}\,=\,\bigcup_{\{k_i\}}\M_{\{k_i\}}.
\eeq
Here $\{k_i\}$ runs over sequences of integers 
satisfying the conditions (\ref{range1}) and 
(\ref{range2}),
and $\M_{\{k_i\}}$ is roughly a direct product 
${\cal N}_{\{k_i\}}\times \HH_{\{k_i\}}$ with its boundary
being included. The space
\beq
\M_{\{0\}}\,=\,Hilb^{[k]}(\C^2/\Z_n)
\eeq
is the unique Coulomb branch of dimension $k$, on which 
the global $SU(n)$ symmetry acts trivially.
Any other branch $\M_{\{k_i\}}$ has a non-trivial action 
of $SU(n)$ and touches the Coulomb branch along the 
submanifold
$\overline{{\cal N}_{\{k_i\}}}$. If $\{k_i\}\leq 
\{k_i^{\prime}\}$, $\HH_{\{k_i\}}$ is embedded in 
$\HH_{\{k_i^{\prime}\}}$
and it is possible that a boundary of ${\cal N}_{\{k_i\}}$ 
intersects with ${\cal N}_{\{k_i^{\prime}\}}$.
When this happens to be the case, a boundary of 
$\M_{\{k_i\}}$ is embedded in $\M_{\{k_i^{\prime}\}}$
as a submanifold. To know whether this really happens or 
not, we need more information on the adjacency relations of 
the ${\cal N}_{\{k_i\}}$'s
in $Hilb^{[k]}(\C^2/\Z_n)$.

\section{Duality for $U(k)$ Gauge Theories III:
 T-Duality and Extremal Transition Picture}

In this section, we discuss how to understand the mirror
symmetry between the A and B-models from the string theory
view point. It has been suggested in \cite{si}, \cite{SS} 
that the mirror symmetry 
in three dimensions should be a consequence of the T-duality between IIA and 
IIB strings. The type IIA string compactified on a Calabi-Yau 3-fold $M$
times $S^1$ is, by the T-duality, equivalent to the type IIB string 
on the same geometry except for the change of the radius of $S^1$. 
Under the T-duality, the vector and the hypermultiplet moduli spaces of 
the two theories are interchanged. This is exactly the situation 
of the mirror symmetry in three dimensions. Here we will examine how
this suggestion can be implemented explicitly in our case. 

There is a particular Calabi-Yau 3-fold $M$ on which the type IIA
string gives the field content of the A-model 
\cite{klemm,kmp,bkkv,KV}. In order to turn off gravity, we take the 
Planck mass to
infinity after the compactification.
At the same time, we would like to have finite masses for relevant charged
particles coming from D-branes wrapping cycles in $M$. 
Thus we have to consider a singular limit of $M$, where we
scale the relevant K\"ahler moduli of $M$ to zero simultaneously. 
In fact a local description of the singularity of $M$ is sufficient in 
order to
understand the field theory limit of the compactified IIA string 
\cite{kkv}. To realize the A-model in three dimensions,  
we send the radius $R_A$ of $S^1$ to zero also while
keeping $R_A m_{string}$ finite. 

This compactification of the IIA string is related, by the T-duality, 
to the type IIB string on $M \times S^1$ with the radius of $S^1$ being $R_B = 
(R_A m_{string}^2)^{-1}$. Since $R_B$ scales as $1/m_{string}$,
the T-dual of the A-model should also give a three-dimensional 
field theory with rigid $N=4$
supersymmetry. In fact, in the case of $k=1$ with $n$ being arbitrary, 
we will show that the type IIB string on $M$ reproduces the field 
content of the B-model. This means that, in this case, the mirror 
symmetry of the A and B-models can indeed be interpreted as 
a consequence of the T-duality of the type IIA and IIB string theories. 
We also present some evidences for the $k >1$ case.

\subsection{A-Model}

The A-model of the gauge theory arises from 
the type IIA string on a Calabi-Yau 3-fold $M$ 
constructed as a family of K3 fibered over a complex one-dimensional 
torus
${\cal C}$ \cite{klemm,kmp}. In order to reproduce the field content 
in the
A-model, namely:

\noindent
$\bullet$ vector multiplet with $U(k)$ gauge group,

\noindent
$\bullet$ one hypermultiplet $(A, \tilde{A})$ in the adjoint
representation,

\noindent
$\bullet$  $n$ hypermultiplets $(Q_i, \tilde{Q}_i)$ ($i=1,...,n$)
in the fundamental representation, 

\noindent
we consider a case
when K3 has singularities of type $A_k$ at $n$ isolated
points $w=w_1,...,w_n$ on ${\cal C}$ which are resolved to type
$A_{k-1}$ over a generic point \cite{bkkv,KV}. 
The geometry of the Calabi-Yau manifold $M$ near the
singularities can be modeled by the equation,
\beq
  z^k  (z+P_n(w)) + x^2 + y^2 = 0,
\label{singular}
\eeq
where $P_n(w)$ is a polynomial 
of degree $n$ with $n$ zeroes at $w_1,...,w_n$,  
and $(x,y,z)$ parametrize the K3 fiber.
We can see that the fiber develops an  $A_k$ singularity
at $n$ points on ${\cal C}$ where $P_n(w)=0$.  The $A_{k-1}$ singularity on 
a generic fiber can be resolved as
\beq
  \prod_{a=1}^k(z+\mu_a) \cdot (z+P_n(w)) + x^2 + y^2 = 0.
\label{resolve}
\eeq
However, for each
$a=1,...,k$, the
fiber still has $A_1$ singularity at $n$ points on ${\cal C}$
solving $P_n(w) = \mu_a$.

Let us demonstrate that this geometry indeed
generates the field content of the A-model. Due to the $A_k$
singularity, there are ${1 \over 2} k(k+1)$ 2-cycles
$S_{ab}$ ($a,b = 1,..., k+1; a < b$) on each fiber. The cycle
$S_{ab}$ with $a,b \leq k$ vanishes when we choose the 
complex moduli $\mu_a = \mu_b$. On the other hand,
$S_{a(k+1)}$ always vanishes
at $n$ points on ${\cal C}$ satisfying $P_n(w) = \mu_a$. 
Among these cycles,
$k$ of them are homologically independent. We can choose 
$S_{12}$, $S_{23}$, ..., $S_{k(k+1)}$ as primitive cycles and 
correspondingly there are
$k$ K\"ahler moduli of $M$. For later convenience, we denote the
one associated to $S_{a(a+1)}$ by $(t_a - t_{a+1})$ with $t_{k+1}=0$
(alternatively one may choose $S_{1(k+1)},..., S_{k(k+1)}$ as
primitive cycles and $t^a$ as the K\"ahler moduli associated to them).
Each of the K\"ahler moduli can be identified as a vev of 
charge neutral scalar component of the vector multiplet.
Since the base of $M$ is a torus ${\cal C}$, there is also 
a generator $\eta_a$ of $H^{2,1}(M)$ associated to
each $S_{a(k+1)}$. To see this explicitly,
one may take the $(1,1)$ form on the fiber corresponding to 
$S_{a(k+1)}$ and tensor it with the holomorphic 1-form on the base 
${\cal C}$.
A 3-cycle dual to $\eta_a$ is $S^1$ of the base ${\cal C}$ times
$S_{a(k+1)}$ of the fiber\footnote{In general, if the base is a
genus-$g$ 
curve, 
each $S_{a(k+1)}$ should give
$g$ elements of $H^{2,1}(M)$ since there are $g$ holomorphic 1-forms.}. 
In fact these $\eta_a \in H^{2,1}(M)$ correspond to the complex modulus
$\mu_a$ in the resolved space (\ref{resolve}).   
The complex modulus $\mu_a$ together with a vev of
the RR 3-form $B^{(3)}$ on $S^1 \times S_{a(k+1)}$ make 
charge neutral scalar components of the adjoint hypermultiplet
$(A, \tilde{A})$. The charged components of the vector and the adjoint
hypermultiplets correspond to wrapping D2-branes on the 2-cycles
$S_{ab}$ ($a,b=1,...,k$). Among them, the cycles $S_{a(a+1)}$
correspond to simple roots of $U(k)$ while others correspond
to non-simple roots. 
 
As we mentioned in the above, for generic values of $\mu_a$,
$S_{ab}$ with $a,b \leq k$ are non-vanishing, but each
$S_{a(k+1)}$ vanishes at $n$ special points satisfying
$P_n(w) = \mu_a$. The cycle $S_{a(k+1)}$
is homologous to the sum of the primitive cycles
$S_{a(a+1)} \cup S_{(a+1)(a+2)} \cup \cdots \cup S_{k(k+1)}$.
Thus, by wrapping a D2-brane on $S_{a(k+1)}$, we find one 
hypermultiplet carrying charges
$q_a=1$ and $q_b=0$ ($b \neq a$). Here $q_a$ means the charge for
the $U(1)$ vector associated to the K\"ahler moduli $t^a$. Of course,
if we choose $\mu_a = 0$ for all $a$, the vanishing of $S_{a(k+1)}$ 
takes place 
at the zeroes of $P_n(w)$. Thus we find that, from 
each of the $n$ exceptional fibers,  we obtain one hypermultiplet 
in the fundamental representation of $U(k)$. 
This completes the field content of the A-model. 

Now let us examine the structure of the Coulomb branch of the
A-model using this string theory construction.  
In four dimensions, the RR 3-form $B^{(3)}$
associated to the 2-cycle $S_{a(k+1)}$ and the RR 5-form $B^{(5)}$ 
associated to its dual 4-cycle $^*S_{a(k+1)}$
give a vector field $v_\mu$ and its dual
$^*v_\mu$.
Upon compactification to three dimensions, their Wilson line expectation
values on $S^1$ make a complex scalar field, which we denote by $u_a$. 
It is then paired with the scalar component of 
the vector multiplet corresponding
to the K\"ahler moduli $t_a$, and the vector multiplet moduli 
space also become a hyperk\"ahler space. Since the
RR charges are quantized, the RR fields $u_a$ are periodically
identified. Thus the vector multiplet moduli space can be viewed as
a family of tori of the RR field $u_a$ fibered over the K\"ahler moduli 
space of $t_a$. 

Let us examine the conifold 
singularity in the moduli space which is generated when the quantum 
corrected size of one of $S_{a(k+1)}$'s vanishes, paying attention to
the 2-dimensional subspace of $(t_a, u_a)$ taking all
other moduli to be of generic value\footnote{We are considering the
quantum corrected size (the one which takes into account worldsheet
instanton effects) since it is the one that is proportional to
the BPS mass of a D-brane wrapped around a 
cycle homologous to $S_{a(k+1)}$.}.  In fact 
there are $n$ homologous 2-cycles (one at each of the $n$ special points) 
whose quantum volumes
vanish simultaneously in this limit. This is a situation in which the 
extremal transition is possible \cite{GMS}. 
Traveling around the singular 
point, the RR fields experience the monodromy transformation
$u_a \rightarrow u_a + n$ \cite{GMV}. This means 
that the moduli space near the conifold point has an orbifold singularity  
${\bf C}^2/{\bf Z}_n$ in the subspace of $(t_a, u_a)$.   

When the two complex moduli
coincide, $\mu_a =\mu_b$,
there appears an additional symmetry which exchanges
$(t_a, u_a)$ and $(t_b, u_b)$.
Thus, in particular when all the complex moduli coincide,
the vector multiplet moduli space becomes $Sym^k(
{\bf C}^2 /{\bf Z}_n)$. This is exactly the structure of the Coulomb
branch of the A-model which we found previously from the mirror symmetry
and the one-loop test. 

After the extremal transition, the $n$ homologous 2-cycles are replaced
by $n$ 3-cycles with 1 homology relation. Thus the origin of
the 2-dimensional subspace $(t_a, u_a)$ of the vector
multiplet moduli space is connected to $2(n-1)$-dimensional 
subspace of the hypermultiplet moduli space, $(n-1)$ of which
correspond to the complex moduli of $M$. We can repeat this 
procedure $k$ times to completely Higgs the vector multiplet. This gives
us $k+k(n-1) = kn$  hypermultiplets (the first
$k$ are the complex moduli $\mu_a$ which are there before the
extremal transition). This correctly reproduces the Coulomb-Higgs
transition discussed in the previous sections. 

\subsection{B-Model}

Here we will consider the same Calabi-Yau manifold $M$, but put
the IIB string on it. As we mentioned before, if the field content
of the B-model is reproduced in this way, the mirror symmetry may
be considered as a consequence of the T-duality of the IIA and IIB 
string theories. 

To begin with, let us consider a case of $k=1$ with $n$ being arbitrary. 
In this case, we have
a family of K3 fibered over ${\cal C}$, with $n$ special points on which 
K3 develops the $A_1$ singularity. Elsewhere K3 is regular in this case.
We would like to show that the type IIB string theory on this geometry
gives the field content of the B-model:

\noindent
$\bullet$ $n$ vector multiplets $v_i$ ($i=1,...,n$) 
of $U(1)$ gauge group,

\noindent
$\bullet$ $n$ hypermultiplets $(B_{i(i+1)}, B_{(i+1)i})$
with charges $q_i = \pm 1$, $q_{i+1}= \mp1$ and $q_{j \neq i, i+1} = 0$,

\noindent
$\bullet$ a hypermultiplet $(Q_0, \tilde{Q}_0 )$ 
with charges $q_1= 1$, $q_2 = \cdots =q_n =0$

\noindent
The geometry as it is has one complex modulus $\mu$ and 
the one K\"ahler modulus $t$ corresponding to the vanishing
$S^2$ on the fiber. As we have seen in the previous subsection,
the type IIA string on this geometry gives the Coulomb branch
of the A-model. Since the T-duality exchanges the vector and
the hypermultiplet moduli spaces, the type IIB string theory
on the same geometry should be in the Higgs branch. To
identify the field content of the B-model, however,
it seems easier to work in the Coulomb branch. This means that we have 
to perform an extremal transition of the geometry. 

Since there are $n$ vanishing 2-cycles at the $n$ special points
and since they are all homologically equivalent, the extremal
transition changes them into $n$ 3-cycles $S^{(i)}$ ($i=1,...,n$) 
with $1$ homology relation $\sum_{i=1}^n S^{(i)} = 0$.  This gives us 
$(n-1)$ complex moduli, in addition to one complex modulus $\mu$
which had been there before the extremal transition. Let us denote 
complex moduli
associated to $S^{(i)}$ by $(\mu^{(i)} - \mu^{(i+1)})$ with
$\mu^{(n+1)} = \mu^{(1)}$. There is a redundancy in this parametrization
corresponding to simultaneous shift of $\mu^{(i)}$'s,
and we fix it by choosing $\mu^{(1)}$ to be equal to the complex
modulus $\mu$.

Now we can identify the field content of the B-model. The $U(1)^n$
vector multiplet comes from the $n$ complex moduli and the $n$
hypermultiplets $(B_{i(i+1)}, B_{(i+1)i})$ are obtained
by wrapping D3-branes on $S^{(i)}$'s. Since the dimensions of
$H_{2,1}$ is $n$, there must be one more 3-cycle which is not
homologous to $S^{(i)}$'s. In fact it is not difficult to identify one.
Before the extremal transition, there is a unique homology 3-cycle
which is the 2-cycle on the fiber times $S^1$ of ${\cal C}$.
Since the extremal transition is a local operation near the
$n$ special points, this 3-cycle should remain after the transition
as far as we choose $S^1$ to be away from these points. In fact,
in our notation, the complex moduli $\mu^{(1)}$ corresponds to
this 3-cycle. By wrapping a D3-brane on this cycle, we obtain one
hypermultiplet $(Q_0,\tilde{Q}_0)$ with charges $q_1=1$, 
$q_2 = \cdots q_n=0$. This completes the field content of the B-model.
As one can see, the $n$ hypermultiplets $(B_{i(i+1)}, B_{(i+1)i})$
are massless at the conifold point where all $S^{(i)}$ are vanishing.
On the other hand, $(Q_0,\tilde{Q}_0)$ is massive even at the conifold
point. This is also consistent with what we know about the B-model, i.e.
the vev of $(Q_0,\tilde{Q}_0)$ is zero in both Coulomb and Higgs branches.
Thus we have found that, in this case, the B-model arises from the type 
IIB string on $M \times S^1 \times {\bf R}^3$.
This shows the mirror symmetry of 
the A and B-models is in fact a consequence of
the T-duality of the IIA and IIB string
theories.

Let us turn to general case  when both $k$ and $n$
are arbitrary. We would like to identity

\noindent
$\bullet$ $n$ vector multiplets $v_i$ ($i=1,...,n$) of $U(k)$ gauge
group,

\noindent
$\bullet$ $n$ hypermultiplets $(B_{i(i+1)}, B_{(i+1)i})$ where $B_{i(i+1)
}$ is in $({\bf k}_i, {\bf k}^{\ast}_{i+1})$ of $U(k)_i \times U(k)_{i+1}$,

\noindent
$\bullet$ one hypermultiplet $(Q_0, \tilde{Q_0})$ in the fundamental
representation of the first $U(k)$.

Before the extremal transition, the number of holomogy 3-cycles is 
$k$, and this corresponds to the number of unbroken $U(1)$ gauge
symmetries in the fully Higgsed branch of the B-model. 
After the extremal transition, the $k$ 2-cycles at each of the $n$
special 
points
are replaced by $k$ 3-cycles $S_a^{(i)}$ ($a=1,...,k$; $i=1,...,n$)
with $k$ homology relations. Thus the number of homology 3-cycles becomes 
$k + (n-1)k = nk$ after the series of transitions, 
and this also agrees with the
number of unbroken $U(1)$ symmetries in the Coulomb branch of the
B-model. By counting charges with respect to these $U(1)$'s, we can
identity wrappings of D3-branes on $S_a^{(i)}$'s as diagonal elements
of $(B_{i(i+1)}, B_{(i+1)i})$, and wrappings of D3-brane on the original
$k$ 3-cycles as $(Q_0, \tilde{Q}_0)$.

We have not yet identified the roots of $U(k)^n$ and the off-diagonal 
elements of $(B_{i(i+1)}, B_{(i+1)i})$. Before the extremal
transitions, in additions to the vanishing 2-cycles at the $n$ special
fibers, there are ${1 \over 2}k(k-1)$ 2-cycles $S_{ab}$. After the
extremal transitions, they should also transform into 3-cycles. In fact,
they appear to carry appropriate $U(1)$ charges to be identified
with these fields. It would be very interesting to work out the
relevant homology relations among the 3-cycles after the extremal
transition and to fully identify the fields in the B-model.

\section{Duality for $Sp(k)$ Gauge Theories}

In this section we study the second proposed family of dualities 
for $Sp(k)$ gauge theories.
We  provide the counting  evidence for this duality proposal,
study  the quantum corrections, 
derive the mirror map and use D-brane probes and the 
Type I - M-theory duality to further
support the gauge theory picture.

\subsection{Counting Evidence}
Again, as a first necessary
evidence for the duality between the A and B models we count 
in quaternionic units 
the dimensions of the Higgs and Coulomb branches and the number of FI and mass
terms. 
\noindent\\
{\bf A-model:} The dimension of the Coulomb branch is the rank of the gauge group
which is $d_V=k$. The dimension of the Higgs branch is the dimension of the
hypermultiplet content ($2nk + 2k^2-k$) minus the dimension of the gauge group
($2k^2 +k$).  
Thus, $d_H = 2k(n-1)$.
The number of FI terms is zero since there are no $U(1)$ factors in the gauge
group,
and the number of mass parameters equals $n+1$.
\noindent\\
{\bf B-model:} The dimension of the Coulomb branch 
 is the rank of $U(k)^4U(2k)^{n-3}$, thus $d_V=2k(n-1)$. 
 The dimension of the Higgs branch is the dimension of the
hypermultiplet content ($k + 4(2k^2) + (n-4)(4k^2)$) 
minus the dimension of the gauge group ($4k^2 +(n-3)4k^2$),
thus $d_H = k$.
The number of FI terms is $n+1$,
while the number of mass parameters
$n_m = (n+1) -(n+1) =0$.
Altogether, we have the following table:
\vskip 0.2cm
\begin{center}
\renewcommand{\arraystretch}{1.5}
\begin{tabular}{|c|c|c|c|c|}     \hline
Model   &  $d_V$ & $d_H$ & $n_{\zeta}$ & $n_m$ \\ \hline
       A  & $k$ & $2k(n-1)$ & $0$ & $n+1$ \\  \hline     
       B &  $2k(n-1)$ & $k$ & $n+1$ & $0$ \\ \hline 
\end{tabular}
\end{center}
\begin{center}
Table 10: The dimension of the Coulomb and Higgs branches
and the number of mass and FI parameters of A and B models
\end{center}

The counting shows that we have the required symmetry under
A-model $\leftrightarrow$ B-model,
$d_V \leftrightarrow d_H$ and $n_{\zeta} \leftrightarrow n_m$.

\subsection{A model - One-loop Corrections}

In this section we compute the one-loop corrections to the metric on the 
Coulomb branch of the A model with $Sp(k)$ gauge group,
one hypermultiplet in the antisymmetric representation and $n$
hypermultiplets in the fundamental representation.

Let us parametrize the scalars that minimize the potential energy
(\ref{V}) of A model by
\beq
\vp = diag[\vr_1,
 -\vr_1,...,\vr_k,-\vr_k]
\comma
\label{vevp}
\eeq
where as before $\vp = (\phi^1,\phi^2,\phi^3)$.

The one-loop corrected metric of the Coulomb branch of A model 
takes the form 
\beqar
g_{aa} &=& \frac{1}{e^2} - \sum_{b\neq a}^{k}\left(\frac{1}{|\vr_a-\vr_b|}
+\frac{1}{|\vr_a+\vr_b|} \right)  -\frac{2}{|\vr_a|} + 
\sum_{i=1}^{n}\left(\frac{1}{2|\vr_a-\vm_i|} 
+ \frac{1}{2|\vr_a+\vm_i|}\right)\non\\
&+&\sum_{b\neq a}^{k}\left(
 \frac{1}{|\vr_a-\vr_b +\vm_{as}|}
+\frac{1}{|\vr_a+\vr_b +\vm_{as}|} + \frac{1}{|\vr_a-\vr_b -\vm_{as}|} +
\frac{1}{|\vr_a+\vr_b -\vm_{as}|}\right) \non\\
g_{ab} &=&\frac{1}{|\vr_a-\vr_b|}+\frac{1}{|\vr_a+\vr_b|} + \frac{2}{|\vr_a|}
+\frac{2}{|\vr_b|}~~~~~~~~~~~~~~~~~~~~~~~a \neq b\non\\
&-& \left(\frac{1}{|\vr_a-\vr_b +\vm_{as}|}+ \frac{1}{|\vr_a+\vr_b +\vm_{as}|}
+ \frac{1}{|\vr_a-\vr_b -\vm_{as}|}+ \frac{1}{|\vr_a+\vr_b -\vm_{as}|}\right)
\stop
\label{onesp}
\eeqar

As for the $U(k)$ gauge group case,
in order to compute the one-loop correction one need only consider 
all possible  one-loop diagrams with two gauge fields
on the external legs and a vector multiplet 
or hypermultiplet running in the
loop.
Reduction in the number of colors $k$ and flavors $n$ imply
that the all coefficients of the different diagrams are independent of $k,n$.
The hyperk\"ahler properties of the metric (\ref{kahler})
 implies that the contributions of the vector multiplet and the antisymmetric
hypermultiplet to the diagonal and off diagonal elements of the metric are
of opposite sign and the same absolute value, and that
 the hypermultiplets in the fundamental can contribute only to the diagonal
 terms of the metric.
 We then make use of the fact that 
 $Sp(1)$ yields the $SU(2)$ case. For $SU(2)$ the antisymmetric
 representation is trivial.
 Taking the number of fundamentals to be zero 
 fixes the
 coefficient of the vector multiplet contribution, while the case of two 
 massless fundamentals fixes the coefficient of the 
 contribution of the 
 fundamental hypermultiplets.
Finally, the coefficient of the antisymmetric hypermultiplet contribution is fixed
by reading from the 
Lagrangian its relation to that of the fundamental hypermultiplets.

Consider the case where the
mass of the antisymmetric hypermultiplet
vanishes and we have $n>1$ massless
fundamentals. In this case the one-loop metric describes the $k$-symmetric
product of  non-resolved ALE spaces of $D_{n}$ type $X_{D_{n}}$
\footnote{For $n=0$ and $n=1$ we get, after including the one loop and
the monopole corrections,  the $k$-symmetric product of 
an Atiyah-Hitchin
space and its simply connected double cover respectively.}
\beq
{\cal M}_V^{One-loop}({\rm A-model},\vm_{as}=0, \vm_{fund}= 0)
  = Sym^k X_{D_{n}}
\stop
\eeq 
The one-loop result is expected to be
exact in these cases since the metric corresponds to 
a product of $k$ copies of
the moduli space for $SU(2)$
where there are no higher loop or
monopole corrections for $n>1$ \cite{sw}. Thus we conclude that
\beq
{\cal M}_V^{Exact}({\rm A-model},\vm_{as}=0, \vm_{fund}= 0)
  = Sym^k X_{D_{n}}
\stop
\label{symm}
\eeq 
This is exactly the Higgs branch of the B-model when all 
the FI parameters are set to zero \cite{kn}.
Consider now the inclusion of masses for the fundamental 
hypermultiplets while still setting the
mass of the antisymmetric hypermultiplet to zero.
This case still corresponds a product of $k$ copies of
the $SU(2)$ case. However
when the fundamental hypermultiplets are massive  
the metric is no longer positive definite and we expect
that there will be monopole corrections. 
The masses of the hypermultiplets resolve the ALE singularities
and we expect
\beq
{\cal M}_V^{Exact}({\rm A-model},\vm_{as}=0, \vm_{fund}\neq 0)
  = Sym^k \widetilde{X}_{D_{n}}
\comma
\label{symmex}
\eeq 
which is in agreement with the Higgs branch of the B-model
\cite{kn} when the weighted sum (trace) of the FI parameters vanishes.
As we said above, the one-loop  metric  
(\ref{onesp})  for massless antisymmetric and massive
fundamentals the metric in not positive definite, which indicates that indeed there are
monopole corrections that contribute and make the metric positive definite.
In this case the metric is similar to (\ref{one})
and the mechanism of resolving the quotient singularities by adjoint mass
there is like the mechanism of resolving the $D_n$ singularities by fundamental
masses: In both cases there are monopole corrections.

\subsection{The Mirror Map}

The mass of the antisymmetric multiplet is expected to
correspond to the resolution of the quotient
singularities of the symmetric product in (\ref{symm}) and (\ref{symmex}).
The weighted sum of the FI parameters in the B-model (\ref{mapas}) resolves these 
singularities \cite{kn}. The reason for the weights can be traced to the equations
defining the hyperk\"ahler quotient construction of the Higgs branch of the B-model \cite{kn}.
The analogue of equations (\ref{BR}) 
and (\ref{BC}) 
contain in our case two types of matrix equations: Those of
 size $k\times k$ that correspond to the $U(k)$ nodes of the quiver diagram in figure 2 and
 those of size $2k\times 2k$ that correspond to the $U(2k)$ nodes.
 The parameter for the resolution of the symmetric product is  \cite{kn}
 \beq
Tr[
 \vec{\zeta}_0{\bf 1}_{k\times k}
 +  \vec{\zeta}_1{\bf 1}_{k\times k}
 + \sum_{l=2}^{n-2} \vec{\zeta}_l{\bf 1}_{2k\times 2k}
 +  \vec{\zeta}_{n-1}{\bf 1}_{k\times k}
 +  \vec{\zeta}_n {\bf 1}_{k\times k}]
\comma
\eeq   
which, using similar argument as in the $U(k)$ case,
we identify up to an overall constant with $\vm_{as}$ in (\ref{mapas}).

In order to derive the mirror map for the masses of 
the hypermultiplets in the fundamental
representation we use the same reasoning that led to (\ref{mzeta}).
In the Higgs branch of the B model we expect singularities whenever
a linear combination of the FI parameters corresponding to a 
positive root of $D_n$ vanishes. In the Coulomb branch of the A-model
we expect, from the one-loop metric (\ref{onesp}), such singularities to
appear when $\vm_i = \pm \vm_j$\footnote{It is worth to note 
that we do not expect a 
singularity when a single mass $\vm_i \rightarrow 0$:
The
role of a single
mass parameter is to deform but not to resolve a singularity. For instance, for the gauge group
$SU(2)$ with one massless hypermultiplet the Coulomb branch is the double cover of 
the Atiyah-Hitchin
manifold which is smooth. When turning a mass term for the hypermultiplet we get a deformation
to the Dancer manifold \cite{Dancer}.}. 
Requiring that these singularities match yield the following identification
\beq
\vec{m}_{i} - \vec{m}_{i+1}=  2\vec{\zeta}_i,~~~~~~~~~i=1,...,n
~~~~~~~~~~~~~\vec{m}_{n-1} +\vec{m}_{n}=  2\vec{\zeta}_n
\comma
\label{mzeta1}
\eeq
where $\vm_1 = \vec{\zeta}_1$.
Equations (\ref{mzeta1}) are consistent with the mirror map (\ref{mapas}).
One can extend the singularity analysis to include the mass of the
antisymmetric hypermultiplet, and recover the complete mirror
map (\ref{mapas}) that way.

As a further consistency check we repeated part of 
the analysis of section 5, namely verifying that 
the change of the dimension of the hypermultiplet moduli space when turning
masses in the A-model matches the change of the dimension of the vector
multiplet  moduli space of the B-model when turning on
the corresponding FI parameters.

\subsection{D-brane Picture}

We have argued in the previous section that in the A-model the masses of the $n$ fundamental
hypermultiplets resolve the $D_n$ singularities in the Coulomb branch while the mass
of the antisymmetric hypermultiplet resolves the quotient singularities associated with the
symmetric product (\ref{symm}).
In this section we show that this scenario is expected from string theory viewpoint.

It has been suggested that D-branes can be used to probe the space-time geometry and 
the background gauge fields \cite{probe,probe1,probe2,g,probe3,sen,probe4,probe5}.
In particular, enhanced gauge symmetry in the space-time theory is reflected in the D-brane 
world volume theory by enhanced global symmetry.

Consider a type I string theory  on $R^7\times T^3$, and $k$ D5-branes wrapping the $T^3$ and
yielding $k$ 2-branes in $R^7$.
When the $k$ branes coincide the world volume theory has 
an $Sp(k)$ gauge group \cite{witten}.
The matter fields consist of 16  hypermultiplets in the fundamental representation
of the gauge group arising from the DN sector and one hypermultiplet in the antisymmetric
representation from the DD sector. 
This is precisely the field content of the A-model of the previous section.
The mass terms for the fundamental hypermultiplets
arise from the Wilson lines around $T^3$. Thus, breaking the $SO(32)$ space-time
gauge group by the Wilson lines corresponds to breaking the $SO(32)$ global symmetry on the
world volume of the brane by masses of the fundamental
hypermultiplets.
For massless hypermultiplets the Higgs branch of the world volume theory
is the moduli space of $SO(32)$ $k$-instantons which is the Higgs branch of the A-model.

Consider now the antisymmetric hypermultiplet $\phi_{as}$. When its mass is zero it can get
a generic vev of the form 
\beqar
\langle \phi_{as} \rangle =  
\left ( \begin{array}{ccccc}
 0 & s_1 &  &  &   \\
         -s_1 & 0 &  &  &  \\
             & & \ddots & &\\
            &  & & 0 & s_k \\
             & & & -s_k & 0
          \end{array}
               \right)_{2k\times 2k} 
\comma
\label{vev}
\eeqar
where $s_i$ consists of four real components.
This vev breaks the gauge group to $Sp(1)^k$ and thus separates the $k$ coinciding 
branes.
The Coulomb branch of each brane can be determined  using the duality between 
M-theory on $R^7\times K_3$ and Type I or heterotic string on $R^7\times T^3$ \cite{s}.
Under this duality the the type I five brane wrapping the $T^3$ is mapped to the M-theory
2-brane whose world volume is $R^3\times \{pt \in K_3\}$, which implies that its
Coulomb branch is $K_3$.  The precise Coulomb branch in our case
is an ALE space of $D_{16}$ type. In order to derive that in this context
one has to keep track of the precise
duality map. 
The Coulomb branch for $k$ separated branes is the product of the Coulomb branches
for each brane modded by the action of the Weyl group which  permutes them.
Consequently, 
we get the $k$-symmetric product of the Coulomb branch of a single brane.
This is consistent with the field theory picture for $\vm_{as} = 0$
(\ref{symm}) and (\ref{symmex}).

In order to have a massive antisymmetric hypermultiplet we need to modify
the stringy scenario, so that
$\vm_{as}$ will arise as a parameter of the string theory picture.
If such a stringy picture exists, and if
the mass of the antisymmetric hypermultiplet is different from
zero, it cannot get a vev.
Thus we see that the $k$ branes cannot be separated and 
we expect that the 
Coulomb branch will become
the Hilbert scheme
of $k$ points on an ALE space of $D_{16}$ type. 
It would be interesting to verify this explicitly in string theory.

\section{Duality for $U(k)^n$ Gauge Theories}
In this section  we 
study the third proposed family of dualities for $U(k)^n$ gauge theories.
We  provide the counting  evidence for this duality proposal
and study the Higgs and mixed branches of the dual theories.
Finally, we briefly discuss the mirror map.

\subsection{Counting Evidence}

First, we can count the dimensions of the Coulomb and Higgs branches, as well as
the number of masses and Fayet-Iliopoulos parameters, as we did previously. 
The moduli space of vacua of the theory 
$(U(k)^n;\{v_i\})$ contains a Coulomb branch,
with unbroken gauge group $U(1)^{nk}$. In contrast to the case $n=1$, for $n>1$
this is a pure
Coulomb branch, and not a mixed branch. The moduli space of vacua also
contains a pure Higgs branch (unless $\sum v_i=1$), that is described by
a hyperk\"ahler quotient. The quaternionic dimension of this hyperk\"ahler 
quotient equals $(nk^2 + \sum kv_i - nk^2) = k \sum v_i=km$. 

The number of mass
parameters equals the number of irreducible representations of the gauge group
($n+\sum v_i$) minus the number of $U(1)$'s in the gauge group ($n$), leading to a total
of $\sum v_i=m$. Finally, the number of FI parameters is equal to the number of $U(1)$
factors in the gauge group, which is $n$. These results can be summarized in the following
table
\vskip 0.2cm
\begin{center}
\renewcommand{\arraystretch}{1.5}
\begin{tabular}{|c|c|c|c|c|}     \hline
Model  & $d_V$ & $d_H$ & $n_{\zeta}$ & $n_m$ \\ \hline
$(U(k)^n;\{v_i\})$ & $nk$ & $mk$ & $n$ & $m$ \\ \hline      
$(U(k)^m;\{w_i\})$ & $mk$ & $nk$ & $m$ & $n$ \\ \hline      
\end{tabular}
\end{center}
\begin{center}
Table 11: The dimension of the Coulomb and Higgs branches
and the number of mass and FI parameters of A and B models
\end{center}
where  $\sum v_i = m$ and $\sum w_i=n$.

Again, the counting shows that  we have a symmetry under A-model 
$\leftrightarrow$ B-model, 
$d_V \leftrightarrow d_H$ and $n_{\zeta} \leftrightarrow n_m$, in
accordance with the duality proposal.

\subsection{Mixed Branches}

As a further check of the conjecture we will now consider some of the mixed Coulomb/Higgs
branches that both theories posses in their moduli space of vacua, restricting our
attention to the case where the masses and FI parameters vanish. Such mixed branches
appear when we restrict the vev's of the scalars that parametrize the Coulomb branch
in such a way that some of the matter fields become massless, and can acquire a nonzero
expectation value. Their expectation values parametrize a hyperk\"ahler quotient,
the group being that piece of the unbroken gauge group under which the massless matter
fields are charged.
The global geometry of such mixed branches can be quite
complicated, as the Coulomb branch can receive quantum corrections, but we expect in
general that the mixed branches have the structure of a fiber bundle whose fiber
is described by a hyperk\"ahler quotient. In any case we will here only count the
dimensions of some of the mixed branches, and not consider their global structure.

When analyzing such mixed branches, it may happen the the hyperk\"ahler quotient
corresponds to a case where the group does not act properly on the
hyperk\"ahler
manifold, and the quotient is singular. 
We will be mainly interested in the case where $G$ acts nowhere properly, so
that part of the gauge group is unbroken and we are dealing with a mixed
branch.
Consider such a case and denote the 
hyperk\"ahler manifold by $M$ and the group by $G$. Since $G$ does not act
properly, at every $p\in M$ there is a nontrivial subgroup $G_p$ of
$G$ that leaves $p$ invariant. The submanifold $M_{G_p}$ of points $q\in M$ such
that $G_q=G_p$ is properly acted upon by the centralizer $Z_{G_p}(G)$ 
of $G_p$ in
$G$ (which is the broken part of the gauge group),
and in addition $M_{G_p}$ is hyperk\"ahler. Therefore we can take the
hyperk\"ahler quotient of $M_{G_p}$ with respect to $Z_{G_p}(G)$, and the
result is one of the smooth strata of the hyperk\"ahler quotient $M/G$ of $M$
with respect to $G$. By varying $G_p$, we obtain in this way all the strata
of $M/G$, and two $G_p$'s related by conjugation give rise to the same stratum.

Let us now analyze the mixed branches of the theory $(U(k)^n;\{v_i\})$. We 
impose constraints on the vev's of the scalars that parametrize the Coulomb
branch in such a way that the vev $a$ appears $n^i(a)$ times in the scalars coming
from the $i$th gauge group in $U(k)^n$. This puts us on a submanifold of dimension
$\sum_{i,a} \delta_{n^i(a)\neq 0}$ of the Coulomb branch, with unbroken
gauge group $\otimes_{i,a} U(n^i(a))$. The number of massless matter fields
on this submanifold equals $\sum_i v_i n^i(0)+\sum_{i,a} n^i(a) n^{i+1}(a)$. 
The Higgs branch ${\cal M}_H$
over this submanifold of the Coulomb branch is given
by a direct product of hyperk\"ahler quotients. If we denote each hyperk\"ahler
quotient by its corresponding quiver diagram, we have explicitly
\be \label{j1}
{\cal M}_H = (\otimes_i U(n^i(0)), \{v_i\} ) \times
\prod_{a\neq 0} ( \otimes_{i} U(n^i(a)),\{0,\ldots,0\})
\ee
Now it will in general happen that the groups in (\ref{j1}) do not act properly.
Then we are in the situation of the previous paragraph, and we have to specify
a broken gauge group to describe a stratum of the hyperk\"ahler quotient.
Although more exotic possibilities are possible, a typical broken gauge group
could be $\otimes_{i,a} U(t^i(a))$, and we will restrict our attention to this
type. The centralizer of this subgroup is $\otimes_{i,a} U(n^i(a)-t^i(a))$,
and one easily sees that the stratum associated to it is
a dense submanifold of 
 the hyperk\"ahler
quotient
\be \label{j2}
{\cal M'}_H = (\otimes_i U(n^i(0)-t^i(0)), \{v_i\} ) \times
\prod_{a\neq 0} ( \otimes_{i} U(n^i(a)-t^i(a)),\{0,\ldots,0\})
\ee
Each of the factors $ ( \otimes_{i} U(n^i(a)-t^i(a)),\{0,\ldots,0\})$ has
the property that the quaternionic dimension of the manifold is smaller than
or equal to then dimension of the group, and therefore the group cannot act properly, unless
the quotient has zero dimension. This
implies that for $a\neq 0$
the only consistent choice is $n^i(a)=t^i(a)$, and that we can forget
this part of ${\cal M'}_H$, leaving us with
\be \label{j3}
{\cal M'}_H = (\otimes_i U(n^i(0)-t^i(0)), \{v_i\} ) 
\ee
By assumption, the group $\otimes_i U(n^i(0)-t^i(0))$ acts properly almost everywhere.
But this Higgs branch already emerges over a bigger submanifold of the Coulomb branch:
if we choose $n^i(0)'=n^i(0)-t^i(0)$ and all other $n^i(a)$ arbitrary, we will still
encounter ${\cal M'}_H$ as Higgs branch.

Ignoring the possibility of different types of broken gauge groups, this leads to the following
picture. If we choose integers $k_i$ in such a way that $\otimes_i U(k_i)$ acts
properly in $(\otimes_i U(k_i);\{v_i\})$, then associated to $\{k_i\}$ is a
mixed branch in the moduli space of vacua, where we restrict the vevs of
$k_i$ of the $k$ scalars coming from the $i$th $U(k)$ to vanish, and keep
the others arbitrary. The dimensions of these mixed branches are
\be \label{j4}
(d_V,d_H)=(\sum_{i=0}^{n-1} (k-k_i), \sum_{i=0}^{n-1} k_i v_i - \frac{1}{2} \sum_{i=0}^{n-1} 
 (k_{i+1}-k_i)^2).
\ee

If our duality conjecture is to be correct, we should be able to find similar mixed 
branches, with $d_V$ and $d_H$ interchanged, in the dual theory 
$(U(k)^m;\{w_i\})$. We do not know precisely which sets of integers
$\{k_i\}$ appear in (\ref{j4}), but the results of section~5.1.2 and
section~5.2.2 suggest that the integers have to obey the ``convexity''
condition
\be \label{convx} 2k_i-k_{i-1}-k_{i+1} \leq v_i.
\ee
We have not yet completely solved the problem of finding a mixed 
branch in the B-model for each solution of (\ref{convx}), but luckily
we can show a correspondence in a large class of examples
which is already remarkable in itself.

As our example, we take the theory $(U(k)^n;\{v_i\})$, with $v_i>0$ for each $i$,
and we impose a requirement on the integers $k_i$ that is stronger than (\ref{convx}),
namely that $2k_i \leq v_i$ for each $i$.
For each such
choice we can indeed find a corresponding mixed branch in $(U(k)^m;\{w_i\})$, as
we now describe.
The $w_i$ satisfy $w_{v_0}=w_{v_0+v_1}=\ldots=1$, and all other $w_i$ vanish.
Mixed branches in the B-model
are given by integers $l_i$, $i=0,\ldots,m-1$. Let us choose
them as follows
\be \label{j5}
l_{v_0+v_1+\ldots+v_{i-1}+p} = k-{\rm min}(k_{i-1}+p,k_i,k_{i+1}+v_i-p), 
\qquad p=0,\ldots,v_i.
\ee
Note that the integers $l_i$ also satisfy the convexity condition (\ref{convx}).
One can compute that 
\be d_V=\sum_{i=0}^{m-1} (k-l_i) =\sum_i (k_i v_i - \frac{1}{2}(k_{i+1}-k_i)^2 )
\ee
and that
\be d_H = \sum_{i=0}^{m-1}(  w_i l_i - \frac{1}{2} (l_{i+1}-l_i)^2 ) =\sum_{i=0}^{n-1} (k-k_i).
\ee
These numbers are indeed respectively equal to $d_H$ and $d_V$ of the dual theory.
The mixed branches that may be relevant to study this
duality from the point of view of T-duality and
extremal transitions in string theory, 
correspond to taking all $k_i$ equal to
a fixed $k_0$, and $l_i=k-k_0$. 
To summarize the results for the mixed branches:
\vskip 0.2cm
\begin{center}
\renewcommand{\arraystretch}{1.5}
\begin{tabular}{|c|c|c|c|}     \hline
Model  & mixed branch & $d_V$ & $d_H$  \\ \hline
$(U(k)^n;\{v_i\})$ & $\{k_i\}$ & $\sum(k-k_i)$ & $\sum( k_i v_i - \frac{1}{2} 
 (k_{i+1}-k_i)^2) $\\ \hline   
$(U(k)^m;\{w_i\})$ & 
$\{ l_i \}$ &
$\sum( k_i v_i - \frac{1}{2}  (k_{i+1}-k_i)^2) $ & $\sum(k-k_i)$
\\ \hline      
\end{tabular}
\end{center}
\begin{center}
Table 12: The dimensions of the Coulomb and Higgs mixed branches
\end{center} 
Let us give one explicit example: The theory $(U(k)^3,\{2,6,2\})$ has a mixed branch
with $k_1=1,k_2=2,k_3=1$, with dimension $(d_V,d_H)=(3k-4,15)$. The dual theory
is $(U(k)^{10},\{1,0,1,0,0,0,0,0,1,0\})$,  and according to (\ref{j5}) the corresponding
mixed branch in the dual theory should have $\{l_i\}=\{k-1,k-1,k-1,k-2,k-2,k-2,k-2,k-2,
k-1,k-1\}$. Indeed, the dimensions of this mixed branch are
$d_V=\sum(k-l_i)=15$ and $d_H=\sum_i(l_i  w_i - \frac{1}{2} (l_{i+1}-l_i)^2)=
3k-4$, in accordance with the duality conjecture. 

The analysis of the mixed branches provides
a highly nontrivial check on the consistency of the proposed duality.
The check might be improved even further if one could demonstrate
a one-to-one correspondence between $k_i$ satisfying (\ref{convx})
and $l_i$ satisfying a similar condition.
Also, it would be very interesting to incorporate masses
and FI parameters in the discussion, and to try to derive the mirror map as in section~5.
Right now, the evidence we have for the mirror map (\ref{3mirr}) is based on
an analysis of the singularities. According to theorem 2.8 in \cite{Nakajima2}, 
singularities in the Higgs branch of the A-model will appear if
$\sum n_l \vec{\zeta}_l=0$, where the $n_l$ are nonnegative integers that satisfy
$\sum_i n_i(n_i-n_{i-1}) \leq 2$ and $n_i \leq v_i$. The singularities in
the Higgs branch appear whenever there is an unbroken gauge group. In the
B-model, one can analyze for which masses one expects a singularity in
the Coulomb branch and the appearance of flat directions for the hypermultiplets. 
The result is a generalization of (\ref{genericA}). Requiring that there
is a one-to-one correspondence between these sets of singularities
in the A-model and B-model, one ends up with the mirror map given in
(\ref{3mirr}).

\section{Discussion}

Quiver diagrams provide a natural framework for the
 study of the mirror phenomena in three dimensional
gauge theories.
In this paper we studied three families of mirror gauge 
theories based on 
unitary and symplectic gauge groups.
All these theories contain matter hypermultiplets in 
various representations.
In the absence of matter there is no Higgs branch but
nevertheless it is still natural to ask whether our results
 can shed light on the
 Coulomb branches of the pure gauge
theories, and in fact they do.

Consider first a $U(k)$ gauge theory without hypermultiplets 
in the fundamental
representation but with one massless adjoint hypermultiplet.
Being an $N=8$ supersymmetric gauge theory,
the Coulomb branch  does not receive quantum corrections
and is therefore the hyperk\"ahler manifold
$Sym^k({\bf C^*}\times {\bf C})$ \footnote{Recall 
that ${\bf R}^3\times {\bf S}^1
 \cong \C^* \times \C$.}.
Upon adding a mass to the adjoint 
the Higgs branch is lifted and
we expect the quotient 
singularities to be resolved
and the moduli space to become the Hilbert scheme 
of $k$ points on $\C^* \times \C$.
However we should be cautious since the fact that 
the non-compact space $\C^* \times \C$ 
is hyperk\"ahler
does not guarantee that $Hilb^{[k]}(\C^* \times \C)$ is 
hyperk\"ahler too.
Keeping this point in mind we decouple the adjoint by 
sending its mass
to infinity.
The structure of the moduli space metric, for which we can
gain some understanding 
from the one-loop calculation (\ref{one}), suggests that the 
limit $\vm_{adj} \rightarrow \infty$
scales the metric in such a way that we probe only
a small open subset
of the Hilbert scheme.
This may well still be the same algebraic 
variety as $Hilb^{[k]}(\C^* \times \C)$ if 
it has a  
 hyperk\"ahler quotient construction, similar to the
 one fundamental hypermultiplet case,
 since we expect such
  a 
 construction to be scale invariant\footnote{There will 
 be
  only one parameter in such a hyperk\"ahler quotient 
  construction which is the mass of the
  adjoint hypermultiplet $\vm_{adj}$
  and we can scale the algebraic equation and absorb 
  the scale of $\vm_{adj}$.}.
 Thus we propose that the moduli space is either    
$Hilb^{[k]}(\C^* \times \C)$ or some subset of it.

The moduli space for an $SU(k)$ gauge theory without 
matter hypermultiplets 
follows from the above since the $U(1)$ and the
 $SU(k)$ parts of the $U(k)$ gauge theory decouple in 
 the absence of matter.
 This indicates
 that the moduli space of the  $SU(k)$ gauge
  theory is $Hilb^{[k]}(\C^* \times \C)$
 modded by $\C^* \times \C$, or some subset of it.
It is curious to note that moduli space of pure
$SU(2)$ $k$-monopoles, 
which have been
proposed in \cite{ch} as the moduli space of pure $SU(k)$ gauge
theory  is an open  subset of this space  \cite{ah}.
It is clear however, that in order to correctly identify 
the moduli space we need
a better understanding  of the quantum and monopole 
corrections.

Consider now an $Sp(k)$ gauge theory without hypermultiplets 
in the fundamental
representation and with one massless antisymmetric hypermultiplet.
The Coulomb branch is the symmetric product of 
Atiyah-Hitchin spaces, each of which
we denote by $X_{AH}$. Upon adding a mass term for the antisymmetric 
hypermultiplet, the Higgs branch is lifted, and
we expect to resolve the quotient singularities of the moduli space
and get the Hilbert scheme of $k$ points on the Atiyah-Hitchin space. 
Again, it is not guaranteed
that this space is hyperk\"ahler and we may need a suitable 
subset of it.
Decoupling the antisymmetric hypermultiplet will scale the 
metric in a similar manner as in the 
$U(k)$ case.

The following table summarizes this discussion:
\vskip 0.2cm
\begin{center}
\renewcommand{\arraystretch}{1.5}
\begin{tabular}{|c|c|}     \hline
Gauge Group  & ${\cal M}_V$ \\ \hline
      $U(k)$ & $Hilb^{[k]}(\C^* \times \C)$  \\  \hline     
       $SU(k)$ &  $Hilb^{[k]}(\C^* \times \C)/ (\C^* \times \C)$  \\ \hline 
              $Sp(k)$ & $Hilb^{[k]} X_{AH}$              \\ \hline 
\end{tabular}
\end{center}
\begin{center}
Table 13: The proposed moduli spaces in the absence of matter
\end{center}

There are several natural directions for future studies.
From a field theory viewpoint it is important to understand 
the role of the monopole corrections to the metric
on the Coulomb branch and in particular the mechanism by
which it
 resolves singularities.
It is also interesting to explore the D-brane wrapping 
mechanism that corresponds to the monopole
corrections.
From a string theory  viewpoint it would be important to
 further
explore the stringy origin of the mirror phenomena and the 
mirror map.
In particular it would be interesting to uncover the role 
played by the moduli space 
of D-branes that exists in the wrapping picture.

The detailed study of the moduli space of vacua exhibits a
rich structure of mixed branches 
and possibly non-trivial RG fixed
points which is worth exploring.
We expect that other dual quiver diagrams exist 
which encode the field data for other families
of mirror gauge theories and it would be interesting to find them.

Besides being a remarkable rich structure in its own right, we believe our results
will be useful in obtaining a better understanding of non-perturbative effects in
type II string compactifications,
the physics of small instantons, monopole corrections
in three dimensions and (possibly non-trivial) IR fixed points.

\section*{Acknowledgements}
We would like to thank 
K.\ Bardak\c{c}i, 
I.\ Grojnowski, A.\ Hanany, K.\ Intriligator,
D.\ Morrison, H.\ Nakajima,
R.\ Plesser,  E.\ Silverstein, A.\ Strominger, C.\ Vafa
and Z.\ Yin for discussions. This work is supported in part by 
NSF grant PHY-951497 and DOE grant DE-AC03-76SF00098. JdB is a fellow of
the Miller Institute for Basic Research in Science.

\end{document}